\documentclass[nonacm, sigconf]{acmart}

\usepackage{booktabs}
\usepackage{graphicx}
\usepackage{xcolor}
\usepackage{soul}
\usepackage{caption}
\usepackage{subcaption}

\newcommand{\add}[1]{\textcolor{black}{#1}}
\newcommand{\edit}[1]{\textcolor{black}{#1}}
\newcommand{\revise}[1]{\textcolor{black}{#1}}

\newcommand{\highlight}[1]{\textcolor{black}{#1}}

\AtBeginDocument{%
  }

\begin{document}

\title[Practices, Challenges, \& Opportunities of \add{AI-Supported} Reporting In Local Journalism]{They Think AI Can Do More Than It Actually Can:\\Practices, Challenges, \& Opportunities of \add{AI-Supported} Reporting\\In Local Journalism} 


\author{Besjon Cifliku}
\email{besjon.cifliku@cais-research.de}
\orcid{0009-0007-5081-9531}
\affiliation{%
  \institution{Center For Advanced Internet Studies (CAIS)}
  \city{Bochum}
  \country{Germany}
}

\author{Hendrik Heuer}
\orcid{0000-0003-1919-9016}
\email{hendrik.heuer@cais-research.de}
\affiliation{%
  \institution{Center For Advanced Internet Studies (CAIS)}
  \city{Bochum}
  \country{Germany}
}
\affiliation{%
  \institution{University of Wuppertal}
  \city{Wuppertal}
  \country{Germany}
}

\renewcommand{\shortauthors}{Cifliku and Heuer}

\begin{abstract}
Declining newspaper revenues prompt local newsrooms to adopt automation to maintain efficiency and keep the community informed. However, current research provides a limited understanding of how local journalists work with digital data and which newsroom processes would benefit most from AI-supported (data) reporting. To bridge this gap, we conducted 21 semi-structured interviews with local journalists in Germany. Our study investigates how local journalists use data and AI (RQ1); the challenges they encounter when interacting with data and AI (RQ2); and the self-perceived opportunities of AI-supported reporting systems through the lens of discursive design (RQ3). Our findings reveal that local journalists do not fully leverage AI's potential to support data-related work. Despite local journalists' limited awareness of AI's capabilities, they are willing to use it to process data and discover stories. Finally, we provide recommendations for improving AI-supported reporting in the context of local news, grounded in the journalists' socio-technical perspective and their imagined AI future capabilities.
\end{abstract}


\maketitle

\textbf{This paper is conditionally accepted for the CHI Conference on Human Factors in Computing Systems (CHI ’26), April 13-17, 2026, Barcelona, Spain (Author Draft)}

\section{Introduction}

The number of local newsrooms in Germany is slowly declining~\cite{rudolfaugstein2024wuestenradar}. The downward trend \highlight{increases} pressure to supply local communities with relevant news~\cite{future-of-media-lokaljournalismus} and might scrutinize the public's ability to remain informed. AI and automation can increase efficiency in news reporting, \add{supporting} multiple processes \add{including} creating, gathering, verifying, and distributing news~\cite{Cools2024PerilsPerceptions}. In this context, AI-supported reporting could equip journalists with \add{the} technical means to process the constantly growing volume of digital data~\cite{diakopoulos2019automating}, translating raw numbers into timely\add{,} newsworthy stories essential for digital journalism. On the other hand, as AI \edit{becomes integrated} into the newsroom, local journalists must adapt their skills to use these new technologies effectively.

Although newsrooms invest in cross-functional teams to experiment with AI, only 5\% of the news workforce holds a technology-based degree, according to a 2017 report from the ICFJ surveying \edit{over} 2,700 journalists~\cite{icfj2017news} (see ~\cite{TechnicallyImpressive}). \add{Studies report low numerical literacy among journalists and, therefore, a hesitation to engage with data~\cite{ScottNumericalLiteracy2003, BorgesRey2020}.} \add{However, } little is known about how non-technical local journalists interact with digital \add{raw} data and the challenges they face, a prerequisite for identifying gaps where AI can support. A recent study~\cite{EderFallingBehindLocalJournalismStruggle} \add{suggests} that Germany's local news ecosystem struggles to innovate during the AI transformation, highlighting the need to understand how journalists can \add{effectively integrate} these systems into their workflows. Without computational~\cite{diakopoulos2024data} and data \add{literacy}~\cite{ShullerDataLiteracy2022}, local journalists might be unable to fully realize the potential of AI \add{or even recognize its limitations~\cite{SinaKordonouriJournalisticAgencyAI2025}}. Despite \add{this}, existing work focuses on AI's opportunities and challenges on a broader level \cite{Cools2024PerilsPerceptions, CoolsFromAutomationToTransformation, DiakopoulosAPReport2024, MollerOneSizeFitsSomeJournalisticRole2025, Schellmann2025_AItoolsForJournalism, thomson2025generativeAIJournalism} and often overlooks the implications in local news. \add{Local media is central to democratic values in local communities~\cite{NeillPassiveJournalists2008} as a primary source of news unreported elsewhere~\cite{SjoevaagNunSustutabilityLocalNews}.} \add{Nonetheless}, there is little focus on data awareness and AI-supported reporting in local journalism in Germany \add{through} an HCI lens. To address this gap and inform the future of local journalism and AI, we explore the research questions: 
\begin{itemize}
    \item \textbf{RQ1}: How are local journalists using data and AI in their workflows?
    \item \textbf{RQ2}: What challenges do local journalists in Germany report when working with data and AI?
    \item \textbf{RQ3}: What opportunities do local journalists perceive in AI-supported reporting \add{when interacting with data} regarding a) \add{imagined future technical capabilities} and b) the self-reported implications for their role? 
\end{itemize}

To answer these questions, we conducted 21 semi-structured interviews with local journalists (10 females) in Germany. We developed and presented videos of two research prototypes during the interviews to elicit practical insights from local journalistic workflows. The prototypes aim to automate online data processing tasks following a computational news discovery approach~\cite{NishalCDN25}. We aimed to explore how AI can make data more accessible to non-technical journalists. We not only examine how journalists engage with AI and \add{data in their reporting}, along with the associated implications, challenges, and opportunities, but also \add{provide a comprehensive} analysis of how automation may reshape the socio-technical landscape of local news practices. \add{Through this work, our goal is to provide the requisite background for future studies to formulate data competency recommendations for publishers and media educators, thereby contributing to the design of a more inclusive AI-supported innovation in local media.}

Our qualitative content analysis \add{reveals} that local journalists do not fully leverage AI for data-related \add{tasks}, relying mainly on it for text support and rarely engaging with digital data processing. Their \add{AI and data} usage patterns are highly influenced by how they perceive their role in the newsroom~\cite{MollerOneSizeFitsSomeJournalisticRole2025}. However, local journalists consider AI helpful \add{for automating} repetitive tasks, empowering them to strengthen in-depth research, unburden\add{ing} them to create more impactful human-centered stories, and assist\add{ing} in fact-checking, content verification, and investigative work. With this work, (1) we provide an in-depth analysis and empirical understanding of \add{the data-related challenges faced by local journalists with limited digital data literacy} which can inform new tools respecting journalistic needs; (2) share insights into potential opportunities and threats of AI-supported reporting for local journalism zooming in on local journalistic socio-technical perspective; and (3) identify several design possibilities that embody journalistic values, including human agency in data verification, collaborative automation workflows, local news context-aware data interpretation, and \add{multi-}angle reporting.

\section{Background and Related Work}

\add{We first outline digital disruption in local media and how journalists interact with data and AI. We then discuss how data and AI complement each other, shaping the roles of journalists. Finally, we contribute to discussions on AI’s potential and limitations, aiming to help journalists become data proficient or use AI meaningfully.}

\subsection{Local (Data) Journalism}
\label{Background}

\add{Local media serve local communities and marginalized groups, focusing on regional or suburban areas~\cite{NeillPassiveJournalists2008}. National broadcasters scale to broader audiences~\cite{Figl2017DataJournalismSmallNewsrooms} but cannot cover all local community events due to audience prioritization.} Studies report on an increasing trend in \textbf{data} journalism~\cite{LocalJournalismUK, Appelgren2014DataJournalism}, particularly in local news. \add{
However,  local newsrooms struggle with data scarcity and resource limitations, lacking staff, training, and innovation capacity \cite{Figl2017DataJournalismSmallNewsrooms}, while national outlets drive data journalism innovation \cite{BorgesRey2020}. \add{Pressured by editorial constraints and shrinking staff, local outlets rely on external experts~\cite{BorgesRey2020}, which undermines their independence and autonomy. }} 

Due to digitalization~\cite{gillmor2004wethmedia} \edit{and the shift} in consumption patterns~\cite{reutersDigitalReport}, the diversity of local media in Germany~\cite{RSF_Germany_2024} has declined. \add{Although Germany maintains high trust in the news and a strong local media presence compared to the US and some other European countries~\cite{reutersDigitalReport},} \add{the number of local newsrooms is shrinking~\cite{RSF_Germany_2024}.} Consequently, the risk of news deserts~\cite{medill2024localnewsdeserts} grows. \add{The newsroom's size reduction is linked to a diminished public capacity to hold local power accountable~\cite{Cohen2011_ComputationalJournalism}. In addition, due to a drop in advertising revenue, local outlets have lost their monopoly on local information~\cite{SjoevaagNunSustutabilityLocalNews}.}

\add{To sustain profitability, Usher argues that newsrooms prioritize speed and quantity over quality selective news~\cite{Usher2014MakingNews}, eroding audience trust. Lack of trust and financial instability~\cite{SjoevaagNunSustutabilityLocalNews} undermine newsrooms' role as community democracy watchdogs ~\cite{Cohen2011_ComputationalJournalism}, further hindering their capacity to invest in data tools. These limitations motivated us to explore the challenges local journalists face in relation to data~(RQ2).}  

AI \add{could} address local newsrooms' concerns. Automation promises to help journalists produce news on demand while processing and managing digital data without extensive technical expertise~\cite{diakopoulos2019automating}. \add{Prior research on AI-supported journalism underexplores the unique role of local data. There are a limited number of studies on the data interactions of local journalists in Germany (e.g.~\cite{StalphDataLocalJournalism}). Examining local journalists’ data and AI practices reveals efficiency gaps in data processing, which could offer insights for HCI to address local data power dynamics and media sustainability.} 

\add{Despite growing demand for data skills, journalists face editorial resistance and fear of data analysis ~\cite{Weinacht2022Datenjournalismus}. Research suggests integrating data journalism and programming into journalistic academic training~\cite{Weinacht2022Datenjournalismus}. These claims underscore the urgent need to systematically analyze local journalists’ data processes and awareness~(RQ1). Another work describes the challenges journalists face when combining multiple data sources~\cite{Kascia23} while comparing how data journalists and data scientists process data.} Although \add{the paper} shares common themes, \add{our research targets data practices of non-technical local news experts}.

\subsection{AI-Supported Reporting: How Local Journalists Can Make Use of Data?}

\add{This paper examines the interplay between data and AI in local reporting and identifies which data workflows AI can support.} \add{We view AI and data as interrelated concepts that mutually shape each other in an evolving relationship~\cite{Jarke2024AlgorithmicRegimes}. Their complementary relationship reflects the hybrid nature of AI-supported journalism. To establish the terminology,} we operationalize the term \textbf{``AI-supported reporting''}. We consider it to be any system that automates data tasks and generates insights \add{from raw digital data} to support journalists in \add{analyzing data}, shaping patterns into narratives, and creating stories from data (\textit{an idea extended from~\cite{diakopoulos2019automating} and ~\cite{dierickx2024datacentric}}). \add{In the context of this work, we consider digital data to be both structured and unstructured content.} \add{Thus}, we regard AI-supported tools as lowering the \add{technical} barriers of doing data-driven work. \add{We explore AI's potential future impact on local journalistic work and its implications for the professional and normative role of local journalists~(RQ3). We further investigate how data literacy and AI affordances shape local data dynamics. This topic is underrepresented in the literature}.

\add{AI-supported reporting offers local newsrooms an opportunity to uphold their position as trustworthy sources. AI enables novice reporters to adopt watchdog~\cite{Cohen2011_ComputationalJournalism} and data journalism, grounded in statistical rigor and evidence-based storytelling \cite{Figl2017DataJournalismSmallNewsrooms}. AI offers journalists diverse ways to conduct research and navigate data. For instance, \textit{``Der Tagesspiegel'' (a German daily newspaper)} used AI to dynamically map Berlin’s swimming pools into an interactive narrative~\cite{Tagesspiegel2025PrivatePoolsBerlin}, demonstrating the distribution of wealth in the city and addressing the neighborhoods' inequality, thereby sparking public discourse.} \add{Prior} work explores AI's disruptive role in reshaping the news industry~\cite{Lewis2025GenAIDisruptiveJournalism} across various news processes~\cite{DhaeseleerAIDivides2025, CoolsFromAutomationToTransformation, Cools2024PerilsPerceptions}. AI \add{can} assist journalists in practical tasks~\cite{DiakopoulosAPReport2024, beckettYaseen2023generating} such as managing data sources, personalizing content~\cite{hagar2019optimizing}, sensemaking~\cite{nishal2024dejargonizingsciencejournalistsgpt4}, or transcribing~\cite{globalAIjournalism2025transcription}. \add{AI can support} investigative journalists in uncovering the truth~\cite{Broussard02112015, VeerbeekFightingFireWithFire, Stray2019, diakopolous-2024-agent-data-reporting, besjonHendrikCHILBW2025, besjonCHI25NewsFutureWorkshop} or in locating and monitoring various online sources~\cite{nishal2024envisizoningapplicationsimplicationsgenerative} to discover stories computationally~\cite{NishalCDN25, diakopoulous2021NewsLead}. AI can even \add{introduce multiple viewpoints in reporting}~\cite{PetridisEtAl2023_AngleKindling} or adapt news as liquid content~\cite{Spencer2025LiquidContent}. \add{However, many journalists remain at an early stage of adoption due to limited expertise and understanding of appropriate tools~\cite{JonesJonesLuger2022} and institutional fragmentation~\cite{Lewis2025GenAIDisruptiveJournalism}.}

\add{Despite growing discourse}, due to resource shortages and a lack of technical expertise, newsrooms remain experimental~\cite{reuters2025future_journalism, Rinehart2022} and only use AI incrementally~\cite{AspenDigital_AI_GrowingRole2025}, without transforming their journalistic practices~\cite{DhaeseleerAIDivides2025}. Investments in AI specialists are either a burden or unsustainable in the long run if there is no real-world applied scenario to justify the costs (PM Daily Mail)~\cite{reuters2025future_journalism}. In this context, Eder and Sjøvaag discuss the limited innovation in German local newsrooms~\cite{EderFallingBehindLocalJournalismStruggle}. They note that local newsrooms express a strong interest in adopting AI. However, ``local media lags'' due to limited resources and conflicting priorities. The study \add{suggests} that newsrooms may \add{be hesitant to} undertake innovation risks due to their capacity and funding limitations~\cite{EderFallingBehindLocalJournalismStruggle}. Surprisingly, nine in ten newsroom leaders think (generative) AI will transform newsrooms~\cite{reuters2025future_journalism}. \add{R}esearch \add{indicates} that the media may exaggerate the AI capabilities, as \add{published} articles \add{frequently} exclude impacted worker's perspective~\cite{Shorey2025Automation}, framing AI as inevitable yet offering no clear guidelines for \add{integrating} it~\cite{MagalhaesSmit2025_LessHypeMoreDrama} in the newsroom. 

\add{The discussion centers on the ethical issues arising from \edit{AI's}} conflict with journalistic values~\add{\cite{komatsu-ai-embody-values-2020, Nishal2025Values}}, \add{the epistemic uncertainty of journalistic roles}~\cite{Munoriyarwa2025GenerativeAI}, and the interdependence of journalistic values and AI capabilities~\cite{TechnicallyImpressive}. \add{Independence, autonomy, accuracy, and accountability in news work affect the inner workings of AI-supported socio-technical systems~\cite{Nishal2025Values}. Indeed, the tension between data, AI, and journalistic skills constitutes a power dynamic~\cite{BorgesRey2020} and a clash of competing priorities that underscores the complexity of navigating the socio-technical friction arising from the news automation~\cite{dodds_zamith_lewis_2025}.} \add{These concerns demand action to ensure that AI conforms to journalists' values~\cite{Nishal2025Values} in support of local media sustainability. Thus, we need to examine current AI practices in local news. Then we can align journalists' AI future visions with feasible and actionable insights, based on how local journalists interact with data and the self-reported challenges they perceive}.

\subsection{AI, Data, and Journalism: Beyond Tensions and Implications}

\edit{The Web} enabled online actors to quickly \edit{produce} massive amount\add{s} of information at scale. Journalists explore this \textit{raw} content to identify insights and details that can lead to stories. They transform raw data into usable information and then translate it into knowledge~\cite{diakopoulos2019automating, ShullerDataLiteracyFramework2019} \edit{to} inform the public. This process is important in a democratic society where decisions increasingly rely on data~\cite{ShullerDataLiteracyFramework2019}. \add{Journalists use data to articulate alternative perspectives and to provide evidence to validate their stories~\cite{BorgesRey2020}. Data can press local government to address critical social issues~\cite{Figl2017DataJournalismSmallNewsrooms}. However, scholars report a low number of journalists with the necessary skills to engage effectively with data despite a growing demand for data expertise~\cite{BorgesRey2020}.} 

According to Schüller, data literacy encompasses the creation of data, collecting it from various sources, and the capacity to manage and critically interpret it to facilitate practical applications~\cite{ShullerDataLiteracyFramework2019}. This definition is consistent with~\cite{DataLiteracyDefinition}.

\add{Scholars acknowledge the lack of data skills~\cite{Weinacht2022Datenjournalismus, Appelgren2014DataJournalism, BorgesRey2020, StalphDataLocalJournalism} and limited AI literacy~\cite{CoolsFromAutomationToTransformation, DhaeseleerAIDivides2025} among journalists (be it local, regional, or national). Prior works rarely focus on the specific constraints that non-technical local journalists encounter when interacting with data and overlook the impact of AI on local data processes. Understanding local journalists' challenges in data interaction is key to explaining how AI-supported reporting can address them. Many studies call for additional data training but provide limited practical support for less technically skilled journalists to cultivate such data competencies. We hope that our contribution helps newsrooms develop data skills for journalists in response to identified data challenges and guides local newsrooms in developing their data workflows. These challenges can serve as incentives for targeted training, journalistic lectures, and workshops. } 

\add{Journalists do not necessarily believe they require advanced data skills~\cite{BorgesRey2020, LocalJournalismUK}. They perceive their profession as centered on storytelling, even though much of journalistic reporting involves numerical data.~\cite{ScottNumbersInNews2002, ScottNumericalLiteracy2003}.} Interestingly, Kõuts-Klemm reports that journalists often receive data in polished formats, from press releases or academic studies~\cite{KoutsKlemmRagne2019}. Experts analyze and interpret these materials, so journalists rarely evaluate the statistics independently~\cite{Appelgren2014DataJournalism, KoutsKlemmRagne2019}. This data reliance diminishes the trust and reliability of local media~\cite{NeillPassiveJournalists2008}. 

\add{Despite journalists being aware of the potential benefits, many choose to avoid the complexity of data work~\cite{ScottNumericalLiteracy2003, BorgesRey2020}. The AI functionalities for querying and combining sources into usable data can improve the overall quality of local data, promoting more data-driven work. } \add{Nonetheless, journalists resist changes and struggle to adapt their skills~\cite{BorgesRey2020}. The skill gap is compounded by the entrenched challenges of adapting to new workflows, particularly in smaller local newsrooms~\cite{BorgesRey2020} where individuals often shoulder multiple roles. Understanding how to interlink local data processes into a single AI-supported workflow can facilitate the adoption of AI among local journalists. } 

\edit{Journalists'} views on \add{data} automation reflect their data-journalistic skills and the perceived value that data conveys~\cite{thaslerKordonouri2025automatedUK}. Some prior work suggests a correlation between journalists’ data experience and their appraisal of data journalism~\cite{Appelgren2014DataJournalism}, while others highlight how AI shapes journalists' professional identities~\cite{MollerOneSizeFitsSomeJournalisticRole2025}. Møller et al. found that journalists with technological roles or who self-identify as data-related specialists hold more realistic views of AI than others who hype its potential~\cite{MollerOneSizeFitsSomeJournalisticRole2025}. \add{Journalists who familiarize themselves with AI have better expectations of what AI can do and its limitations~\cite{SinaKordonouriJournalisticAgencyAI2025}.} \add{Investigating how such patterns manifest among non-technical local journalists could inform more effective strategies for strengthening journalists' data skills.} Although not on a local level, these studies resonate with our findings as many journalists consider \add{AI} tools helpful for improving their work, and have concerns about the potential impacts on their autonomy. 

\add{While AI support may influence journalists’ confidence when working with data, there is insufficient literature to support this claim. On one hand, r}esearch shows that \add{working with data} \add{might} improve data skills among journalists and increase news variety~\cite{thaslerKordonouri2025automatedUK, Appelgren2014DataJournalism}. On the other hand, AI-supported reporting raises tension between its practicality and hype~\cite{dodds_zamith_lewis_2025, MagalhaesSmit2025_LessHypeMoreDrama}, requiring journalists to constantly negotiate and balance agency~\cite{SinaKordonouriJournalisticAgencyAI2025} and dependency~\cite{dodds_zamith_lewis_2025} without mindlessly trusting AI developers~\cite{CoolsKoliska2024}. \add{Research has the potential to democratize data access for journalists regardless of their technical expertise. However, this tension highlights the need to understand how AI and data shape current local journalistic practices, as well as journalists' beliefs on the impact of future AI-supported reporting on journalism~(RQ3).} 

Jarke and Breiter critically argue (based on \cite{JonesMcCoy3029}) that even though automation can help detect patterns in data, these patterns are meaningless if they are not properly contextualized and interpreted, and can even become harmful if put in the wrong context~\cite{JarkeReport19}. If journalists do not interpret the information correctly, the value it conveys gets lost in reports~\cite{KoutsKlemmRagne2019}. Similarly, Verran highlights that although numbers appear to be simple values, they can hide answers to complex moral and political stories which journalists need to untangle~\cite{verran2021narrating}. \add{To address these challenges, we argue that journalists must develop at least a basic level of data awareness and the skills to process data independently~\cite{LocalJournalismUK}, reducing the risks of misinterpretation and reliance on other parties~\cite{BorgesRey2020} that can influence their autonomy. On the topic of journalistic reflexivity, Borges-Rey suggests that equipping traditional journalists with data skills, basic-level programming, and a fundamental understanding of web dynamics is more rewarding than instructing programmers to do complex journalism proficiently~\cite{BorgesRey2020}.} Our research concentrates on the relationship \add{among} non-technical local journalists, data, and AI, \add{addressing the existing knowledge gap in this area.} \add{This study} investigate\add{s} how AI can impact the \edit{\textit{journalist-data}} relationship, proposing that AI-\edit{supported} reporting can alleviate some concerns local journalists have about using data.  

In sum, to develop suitable tools for journalists, scholars must first understand how local journalists interact with data and interpret statistics. Research must \add{identify} the hurdles that prevent local journalists from interacting with data and \add{propose novel} ways AI can facilitate these complex processes.

\section{Method}

In this work, we dive deeper into how local journalists interact with digital data and AI tools~\textbf{(RQ1)} and investigate the challenges they face when working with multiple data sources and digital data~\textbf{(RQ2)}. Our study is inspired by a \revise{discursive design approach} \revise{\cite{Tharp2019DiscursiveDesign}}, combined with visual elicitation (See ~\cite{johnson2001elicitation}). To elicit journalists' perceptions \add{of} AI-supported reporting and introduce them to \add{the potential of} AI and automation for local news reporting, we used videos of two research prototypes (\add{described} in Section \ref{Prototypes}) combining information retrieval, automated workflows, and \add{digital data processing}. \revise{By adopting a research-through-design approach~\cite{Bardzell2015ResearchThroughDesign, William2012ResearchThroughDesign, Zimmermann2007ResearchThroughDesign, Zimmerman2010ResearchThroughDesign, stappers_giaccardi_2014ResearchThroughDesign}, we surface journalists’ anticipation of future opportunities, threats, and implications of incorporating AI-supported reporting tools into local journalism practices~\textbf{(RQ3)}}. 

\highlight{We conducted this study} in Germany, a country with high press freedom. Although prior studies cover the German \add{data} journalism domain \edit{\cite{StalphDataLocalJournalism, Weinacht2022Datenjournalismus}}, we \add{are among the first to} \add{investigate the overlap of} AI, data\add{,} and local journalism in \add{the} HCI context \add{as perceived by non-technical local journalists}. \add{To guide our investigation}, we conducted semi-structured interviews and followed \textbf{qualitative content analysis}~\cite{Mayring1991}. 

\subsection{Procedure}
The following explains how we recruited participants, our questionnaires, and presents an overview \add{of} the prototypes we \add{showed} during interviews.

\subsubsection{Recruitment}

We applied a purposive systematic sampling strategy~\cite{PurposiveSampling} by recruiting journalists from local and regional newspapers within a 100 km \add{radius of} our location in \add{western} Germany. We considered only online newspapers that updated \add{their} RSS feeds daily and started collecting articles shared via RSS between May and June 2025. \highlight{News is typically categorized into two types: soft and hard news~\cite{HarcupWhatIsNews2017}. We focused only on hard news~\cite{PattersonHardNews2000, ReinemannHardNews2012} (i.e., stories about politics, health, education, economics, and other data-related themes central to informing public opinion).} \highlight{Consequently, we excluded articles related to sports, showbiz, or entertainment. Showbiz articles contain little data as the focus lies on people and celebrities. Although sports rely on data, local sports journalism in Germany often emphasizes entertainment and audience engagement over direct social issues or policy debates. This focus enabled us to analyze how data can practically support the societal impact of local journalism.}

\add{Additionally,} local digital newspapers sometimes feature articles authored by journalists working in other regions in Germany, outside the newspaper's primary coverage area. To ensure the sample validity, we systematically verified the authors' details and excluded articles from national and international news providers, unrelated to local news. We extracted the relevant author details from the RSS feeds and collected their emails when listed on their newspaper profiles. We invited each participant directly via email, advertising our study.

We conducted interviews between June and August 2025. Participants received 25 EUR as \add{an} incentive for joining our study. Nine participants chose to donate their vouchers to the non-profit organization ``Reporters Without Borders''.

\subsubsection{Interview Process}

Prior to each interview, we obtained informed consent from all participants regarding data processing in accordance with GDPR compliance. At the beginning of each interview, we asked for participants' consent to record the meeting. We orally briefed participants that the interview aimed to understand how local journalists without technical expertise work with data, use AI, and the challenges they encounter. \add{We did not define AI-supported journalism during the interviews and left it open-ended to avoid contradicting the local journalists' perspectives on what it entails for them and local news practices.} 

We \add{began} the interview by asking about their demographics, \add{journalistic} experience, their role in the newsroom, and their programming knowledge. We invited them to share how they currently use AI in their work, \add{its impact on their work}, and what they hope for from a future AI assistant capable of handling any repetitive or time-consuming journalistic tasks. Our interviews included questions about participants' experience when working with data \add{(including raw digital data and data sources)}, as well as the challenges they had encountered in the past. Following this, we instructed participants in a scenario-based task, in which they had to investigate crime statistics in the local regions. This task helped us to elicit difficulties journalists would face when working with \add{digital} data. The task included collecting and analyzing online police reports from hundreds of official online press releases. We asked participants how they would begin reporting on such a story, how they would gather and analyze data without \add{the aid of} AI, and then how they would approach the same scenario with AI support. 

Afterwards, we showed the participants the prototype videos, followed by questions on their perceptions and thoughts. \add{We discussed with the participants the perceived opportunities and threats associated with the presented approaches, while considering the strengths and weaknesses that impact the future outlook on AI-supported journalistic tools. We repeatedly questioned participants to reconsider their statements and explore alternative interpretations or assumptions, prompting critical reflection. } We wrapped each interview by asking participants \add{the} plausibility of using AI-supported reporting in their daily work on a Likert scale, followed by questions about the rationale behind their choice. \add{The purpose of the Likert scale was to assess journalists' openness to such technologies in a more structured manner, thereby adding nuance to our qualitative coding.} We further explored journalists’ concerns about using AI for data work, considering the potential role of HCI researchers in supporting them, identifying tools that could best serve local reporting, and assessing their interest in testing our prototypes and engaging in longer-term collaborations. We concluded the session with an open discussion about \add{the future implications of} AI and data in local news.

\subsection{Research Prototypes}
\label{Prototypes}
\begin{figure*}[p]
\centering
\caption{This figure presents our two operational prototypes: Figure \ref{fig:prototypeA}a, Automating-Through-Demonstration, appears in panels 1–3, whereas Figure \ref{fig:prototypeB}b, Automating-Through-Words, appears in panels 4–6. We did not include the city crime map scenario to preserve our anonymity \revise{during reviews}, since it \add{would have} disclosed our location. In this scenario, the user wants to select all football teams listed on Wikipedia and identify their stadiums. The user can select all team names simultaneously by interacting with the table on the Wikipedia page. They can select all entries in the table \highlight{in a single step}. The system records these actions graphically and opens the first result/link. The user then selects the required text (2.). After demonstrating this once, the system can replay the actions continuously until it retrieves all the initially selected elements. The system generates an automation script in Typescript as shown in step (3.). The user can inspect and edit the script. The prototype shown in Figure~\ref{fig:prototypeB} at step (4.) is our second, an automated browser that accepts user instructions through prompts. The crime map in step (6.) serves as a template that illustrates the information presented during interviews generated automatically after analyzing the data. In both prototypes, the user can examine the system logs and download the collected results as a JSON or an Excel File (5.).}
\Description{The figure presents six snippets illustrating our two prototypes. The screenshots appear into two column, and you can view them in a top-to-bottom, left-to-right sequence. The first column comprises three screenshots (positions 1–3) present Prototype A: Automating Through Demonstrations, whereas the final three (positions 5–6) present Prototype B: Automating Through Words. The user selects multiple football team names from a table with football teams in Wikipedia shown in Prototype A. The system records these actions graphically (every click and action you do on the web) and automatically opens the first link of the corresponding team (image 2). The user then highlights the text such as the team's name, team's stadium or owner in the newly opened page (image 3). With this single demonstration, the system can replay the actions across all selected items and generate an automation script. In Prototype B, the user interacts with a crime map template to illustrate interview content. The interface accepts natural-language prompts, as shown in step 4 and 6. Once the user inputs a command on the chat interface, the automated browser system executes tasks based on these user's commands. We did not present the city crime map scenario to preserve our anonymity, since it will disclose our location. The user can inspect execution logs and download the collected results (image 4).}

\begin{minipage}[b]{0.45\textwidth}
    \centering
    \caption*{(a) Prototype 1. Automating-Through-Demonstrations}
    \includegraphics[width=\linewidth]{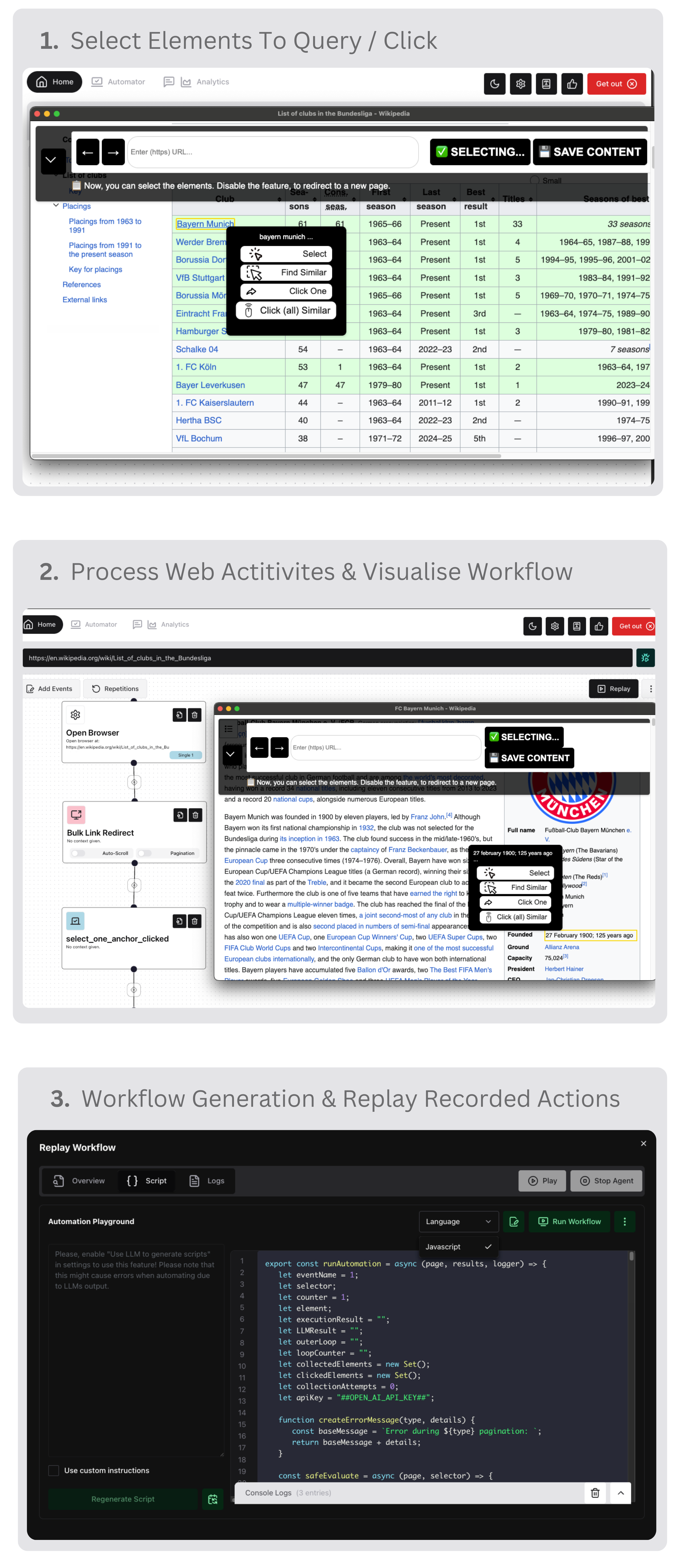}
    \Description{The first prototype automating through demonstrations split into 3 screenshots stacked into a column on the left.}
    \label{fig:prototypeA}
\end{minipage}\hfill
\begin{minipage}[b]{0.45\textwidth}
    \centering
    \caption*{(b) Prototype 2. Automating-Through-Words}
    \includegraphics[width=\linewidth]{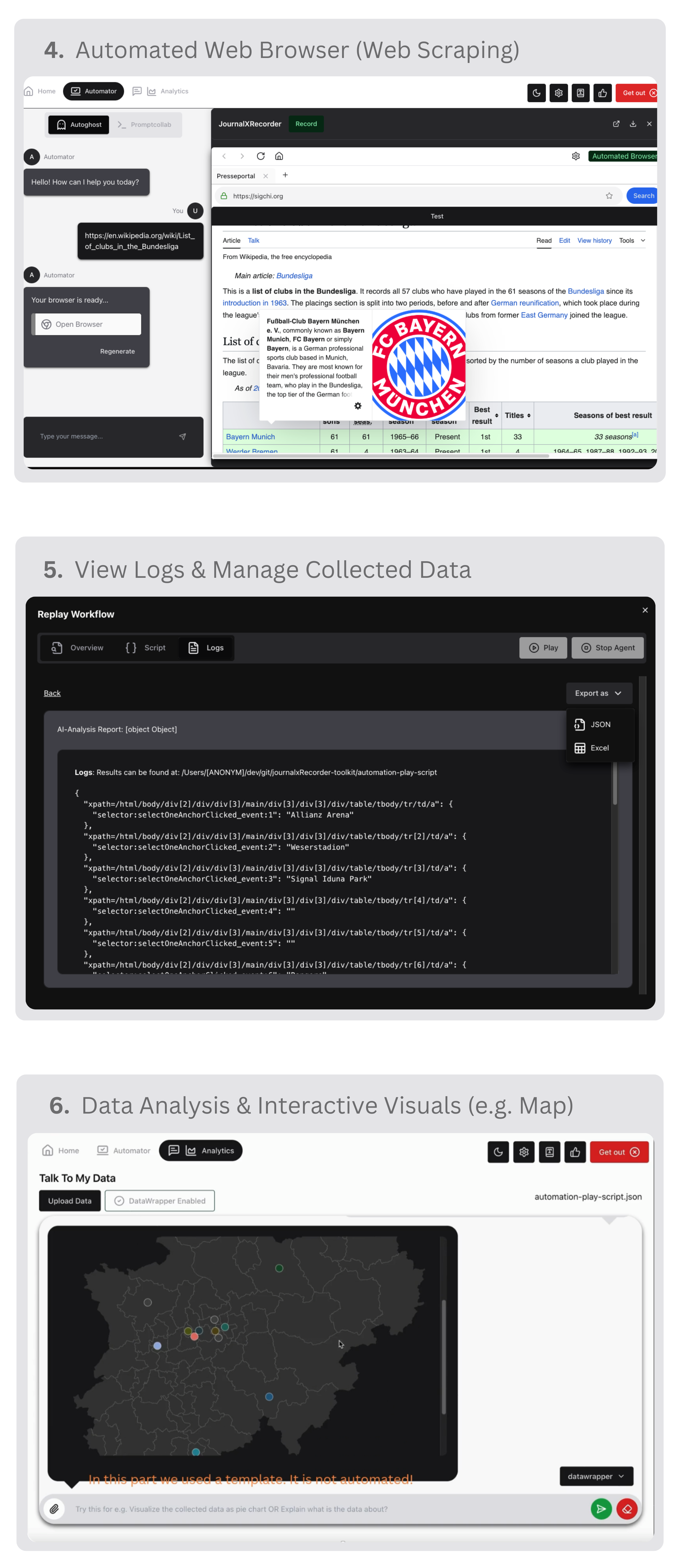}
    \Description{The second prototype automating through words split into 3 screenshots stacked into a column on the right.}
    \label{fig:prototypeB}
\end{minipage}
\label{fig:prototypes}
\end{figure*}

To elicit users' perceptions o\add{f} AI-supported reporting, we developed two research prototypes that employ two different automation techniques (\add{namely,} \textit{automating-through-demonstrations} and \textit{automating-through-words}) to automate online data gathering, \add{analysis,} and reporting. Both prototypes are displayed in Figure~\ref{fig:prototypes}. We created an automated crime reporting scenario and recorded it utilizing both prototypes. In the scenario, the prototypes collected and scraped press releases from online police reports and automatically structured the data in Excel files. The tools then generated journalistic reports and created visual charts and maps highlighting crime locations. In \edit{each} interview, we explained and encouraged that journalists could create other scenarios \add{with} the prototypes related to their repetitive daily tasks. \revise{We included the sports example in Figure \ref{fig:prototypes} to preserve anonymity during review, since our use cases are highly specific to our stakeholders. The example also illustrates that the approach generalizes to other domains, including sports reporting, which we leave for future work.} \add{We decided to use video elicitation prototypes as a sensitizing alternative to direct user interactions. Training journalists to use the prototypes directly would have been time-consuming and could have led to several local journalists not participating. In this way, we avoided the setup overhead during interviews and the security limitations on the user side associated with software installation permissions, user consent, or access to sensitive directories. The scripted videos enabled us to demonstrate a complex user workflow. Therefore, the videos facilitated participants' shared understanding of AI-supported reporting, including the selection, collection, and analysis of data to draw insights. The videos of both prototypes are available in the supplementary materials.}
\\\\
\textbf{Prototype A (Automating Through Demonstrations):} We \revise{applied} the Programming-by-Demonstrations~\cite{ProgrammingByDemonstration} \revise{approach} to enable users to create complex automation scripts by demonstrating tasks, without requiring programming knowledge. The system records users' activity on a web interface (such as clicking, typing, or scrolling)~\cite{besjonHendrikCHILBW2025}. It allows users to revise and repeat these actions several times, potentially extending the script's scope from data collection to other online activities. We used graphs to represent actions in a visual interface. Each node represents an online user activity, which can be modified by users (e.g., by adding more website sources or doing custom web actions). Once users open the website, they can select elements from which they wish to extract data. We presented this approach during interviews as a way to \textit{``teach an assistant''} and show how to automate tasks. This prototype requires human input to record the initial automation blocks, constraining the interactions to \add{specific elements} of the web. The system is autonomous but constrained by human oversight and a clear automation structure, which users can edit and verify. 
\\\\
\textbf{Prototype B (Automating Through Words)}: The prototype resembles a web browser with an attached chat interface, which can autonomously take actions on the web based on user requests. We used prompt engineering to \add{develop} an agent \add{that} automate\add{s} web-based tasks. The agent scraped website content, created custom structured reports, and translated the findings into visual graphics. The system identifies web elements and determines the optimal actions to execute. We presented this approach as a way to \textit{``instruct an assistant''} and tell what to automate. This prototype requires careful prompting and proper context when automating. The system is highly autonomous, self-determining the interactions with web elements. 

\subsection{Participants}

We recruited 21 journalists. Our sample was gender-balanced, with 10 female\add{s}  and 11 male participants. The average age was 42.3 (SD~=~10.8). The participants were highly educated\add{;} all except one \add{held} a Bachelor's degree, three out of four had a Master's, and one had a PhD. The sample is highly skilled, with an average of 19.24 years of work-related experience (SD~=~12.15). In Table~\ref{fig:participants}, we present an overview of our participants’ demographics, including their focus on news reporting. Participants reported on a wide array of subjects, including local news, health, education, economics, and politics. A subset of the participants focused on work related to online news desks. 

\begin{table}[t]
\centering
\caption{Overview of study participants, including years of professional experience, gender, highest completed degree, and current news reporting role. The sample comprises 21 journalists with varied levels of experience and diverse backgrounds.}
\label{fig:participants}
\Description{The table provides demographic and professional background information for 21 participating journalists. Each row represents one participant and includes ID, years of professional experience, gender, highest completed degree, and current news reporting role. Participants’ professional experience ranged from less than 10 years to over 30 years. Six participants had 20+ years of experience, three had 30+ years, and the remainder had 0–10 or 10+ years. In terms of gender, the sample included 10 women and 11 men. For education, most participants held a master’s degree, with others reporting bachelor’s degrees, one with a PhD, and one with a high school diploma. Participants held diverse news roles, spanning local news, online and SEO desks, science reporting, political reporting, education reporting, health reporting, and international politics. Several participants combined multiple roles, such as local news with online desk or news desk responsibilities. Overall, the table highlights a varied group of journalists in terms of professional experience, educational attainment, gender, and reporting specialization, reflecting a broad cross-section of contemporary news practice.}
\begin{tabular}{rllll}
\toprule
\multicolumn{1}{l}{\textbf{ID}} & \multicolumn{1}{c}{\textbf{Exp.}} & \multicolumn{1}{c}{\textbf{Gender}} & \multicolumn{1}{c}{\textbf{Education}} & \multicolumn{1}{l}{\textbf{News Role}} \\ \midrule
P01                                  &  20+&     f                                 &   High-school                                      &  Local News                                       \\ 
P02                                  &  10+&     m                                 &   Masters                                      &    Online Desk (SEO)                                     \\ 
P03                                  &  10+&         m                             &   Masters                                      &   Local News                                      \\ 
P04                                  &  0-10&  f                                    &   Masters                                      &   Local News                                       \\ 
P05                                  &  0-10&   m                                   &   Masters                                      & Local News, Online Desk                                        \\ 
P06                                  &  20+&         m                             &   Bachelors                                      &  Local News, Online Desk                                       \\ 
P07                                  &  10+&  f                                    &  Masters                                       &    Science Reporter                                    \\ 
P08                                  &  10+&  m                                    &  Masters                                       &  Political Reporter                                      \\ 
P09                                  &  20+&        f                              &  Bachelors                                       &     Education Reporter                                   \\ 
P10                                  &   30+&       f                               &   Masters                                      &    Local News                                    \\ 
P11                                  &   30+&    f                                  &   Masters                                      &  Health Reporter                                       \\ 
P12                                  &   10+&     m                                 &   Bachelors                                     &  Local news                                       \\ 
P13                                  &   0-10&  m                                    &    Masters                                     &  Local News, Online Desk                                       \\ 
P14                                  &   30+&         f                             &    Masters                                     &   Education Reporter                                      \\ 
P15                                  &  10+&     m                                 &    Masters                                     &   Local News, Community                                     \\ 
P16                                  & 10+&  f                                    &     Masters                                     &  International Politics                                       \\ 
P17                                  &  30+&      m                                &         Master                                &     Local News                                     \\ 
P18                                  &   10+&           m                           &    Masters                                     &   Local Editor                                   \\ 
P19                                  &    20+&   f                                   &   PhD                                      &   Political Reporter                                     \\ 
P20                                  &    10+&     m                                 &  Bachelors                                       &    Local News                                    \\ 
P21                                  &    20+&     f                                 &  Masters                                       &    Health Reporter                           \\
\bottomrule
\end{tabular}
\end{table}

\subsection{Data Analysis}

The first author conducted the interviews remotely via Webex, except for P02, who preferred to join \add{in person at} our lab. The interviews were scheduled for 45 minutes (Average~=~47:12 min, SD~=~8:30 min). We conducted 14 interviews in English and 7 in German. In some of the German interviews (P01, P06, P09, P11), another researcher from the lab was present to support the first author.  All interviews, except P11, were recorded and automatically transcribed verbatim for in-depth analysis using Webex and F4X. In the case of P11, who did not consent to \add{being recorded during} the interview, two researchers from our team took extensive notes during the session. They merged the notes afterwards to check for misalignment and discussed the interview's validity. We translated the German transcripts into English before the analysis. The first author manually checked all automatic transcripts for any mistakes. 

We applied qualitative content analysis following~\cite{Mayring1991} and openly coded the transcripts in MaxQDA. We followed reflexive~\cite{ReflexiveContentAnalysis} and axial coding~\cite{simmons2017axial} guidelines as we sought a deep understanding of journalistic practices using data. Initially, we applied inductive coding, identifying codes in the transcripts as we thoroughly analyzed each participant's statements. Both authors coded one interview together to arrange an initial code base. We mapped the participant statements to categories of strengths, weaknesses, threats, and opportunities, following the SWOT~\cite{kniazeva2023swot} approach. \add{In the next step,} we \add{categorized} AI-related automation news into four themes based on~\cite{Cools2024PerilsPerceptions}: news discovery, production, verification, and distribution. Further categories \add{surfaced regarding} journalistic values, data workflows, AI limitations, and journalists' perceptions of AI-supported reporting. The first author then analyzed the remaining transcripts, during which new codes emerged. Both authors discussed these codes in weekly meetings, where they iteratively categorized, merged, and divided them, going back and forth in the transcripts to ensure \add{that} participants' statements were clearly labeled. As we reiterated through the transcripts, we distributed the codes into other themes and subcategories related to daily journalistic tasks until we reached an agreement. The discussions helped us to better map the emerging concepts to our research questions.

\section{Results}


We first analyze how journalists interact with data and the tasks they employ AI for~\textbf{(RQ1)}. We then identify the principal challenges \add{non-technical} journalists encounter when \add{interacting} with \add{digital} data and the limitations inherent in their AI tools~\textbf{(RQ2)}. These findings \add{inform} our research on the potential of AI-supported, \add{data-oriented} reporting to support local journalistic practice. Finally, we address journalists’ views on the implications of AI-supported technologies for their self-perceived roles and the future opportunities they envision~\textbf{(RQ3)}. 

\add{As discussed in Section~\ref{Background}, we explore the perspectives of local journalists on data and AI viewed from a non-technical standpoint and identify how these concepts overlap in what we frame as AI-supported reporting. \edit{Most of the reporters interviewed were unfamiliar with digital data practices. They reported limited prior use of AI tools to process digital data.} Understanding the perspectives of non-technical journalists on using AI when working with data helps bridge these knowledge gaps and establishes the theoretical foundation needed to improve journalists' data awareness. The findings from \textbf{RQ1} and \textbf{RQ2} provide a necessary basis for understanding current local practices and for guiding how we, as HCI researchers, can prompt newsrooms to create a \textit{``safe space''}~(P07) that encourages journalists to engage in responsible data work supported by AI.}

\subsection{How are Local Journalists Using Data and AI?~\textbf{(RQ1)}}

\add{This section captures the current state of the local media regarding data and AI integration. Our analysis reveals varying levels of data experience in local media, \add{ranging from no digital data engagement (e.g., P11)} to \add{developing data workflows (e.g, P02)}.} Most participants, though aware of AI's potential in reporting, \add{were unfamiliar or inexperienced with AI-supported applications for digital data processing.} \add{However}, participants \add{viewed} AI as a helpful assistant \add{that} could take \textit{``work off their hands''} (P18). 

\subsubsection{Working with Data} \add{Participants emphasized that data is a core source for local journalism (P08, P09), essential \textit{``for developing good and interesting journalistic stories'}' (P10). They used data and statistics to investigate social issues related to politics, health, education, or crime (P08, P09)}. All participants recognized the \add{difficulties} of processing \add{digital} data~(RQ2). Many did not view \textit{``working with data''} \add{as central to} their core journalistic role (P01, P03, P04, P05, P08, P12, P15, P16, P17), often \add{identifying themselves} as not \textit{``data person''}~(P17) or ``\textit{data journalist} (P07, P15, P16, P18)''. They \add{found data} intimidating and \add{hard} to interpret (P15) on their own, sometimes even ``freaking out'' (P21) those in local news. P16 described her work as not {``data-driven''}, as she focused on text-based storytelling, while P01 highlighted the limited data availability in her daily \add{tasks}. These perspectives \add{reflect limited engagement or interest in} digital data processing, as many journalists \add{have not previously} worked with \add{online} data sources. However, their reporting is grounded in data work. 

\add{Data work is a collaborative effort. For instance, P21 relied on her colleagues skilled in data to process and investigate reports from regional hospitals. Participants described the tedious nature of data tasks as ``boring''~(P21) or ``time-consuming''~(P03, P05, P16). Instead, they emphasized the value of direct community engagement to \textit{"talk to people"} and gather public opinion (P21). This interaction is profound in local news, where data comes from interviews, opinions, and human sources that journalists can ``ask'' (P03, P04, P12, P15, P18, P17). In general, participants relied on direct contacts, government proceedings, or press releases, with limited practical exposure to collecting and processing digital data.}

\subsubsection{Working With AI} Editing and writing \add{remain} core journalistic activities (P05, P07, P08, P09). Participants used AI tools for \textbf{brainstorming} (P04, P05, P07, P08, P10, P15, P16, P20),  \textbf{ideating topics} (P06, P10, P13, P17, P20) and \textbf{summarizing}  (P01, P03, P05, P06, P10, P11, P12, P15, P16, P18). SEO was a recurring topic (P09, P15, P20), with AI used to optimize headlines (P07, P08, P10, P15, P18), and track article performance (P02, P05). AI summaries can make online articles engaging and spark readers' interest~(P01, P04, P05, P12, P15, P18), \add{although participants find that most contain factual errors that require time to resolve (P13, P14)}. P15 argued that AI captures statistical trends on audience preferences but often misses quirky details that make stories memorable (P12). Other applications included \add{\textbf{text generation, editing, proofreading, and republishing}}. For example, P12 used AI to generate headlines:
\begin{quote}
\add{The headlines are really good, and my colleagues like them too. They say ``you are so good at headlines'', but it is not me, it is ChatGPT because it is good at that!}
\end{quote} 
Other emerging themes include transcriptions (P11, P13, P14, P16, P18, P21), translation (P07, P16) as well as document labeling and clustering (P02, P07, P16). Participants leveraged AI as a ``sparring partner'' (P15) to review their work, refine writing and language analysis (P18), or to question ``political commentary' \add{for argument completeness} (P19).  Participants used AI for research in unfamiliar domains  (P03, P06, P07, P16, P17, P19): \textbf{finding sources or experts} (P07, P16, P19), \textbf{exploring leads} (P02, P07, P17), and \textbf{indexing documents and online materials} (P02). P17 emphasized he only used AI at the start of the investigations to generate impressions, followed by ``asking further questions to human beings''. Some newsrooms experimented with print layout automation (P01, P15, P18), AI illustrations (P12) and newsletter automation (P05, P13, P15). \add{AI processes did not interfere with journalistic agency or their creative process. Participants generally did not trust AI-generated responses. Therefore, AI adoption remains cautious, limited, and slow. }

\subsubsection{AI-Supported Reporting \& Local News Innovation}

Newsrooms were only beginning to adopt internal AI systems for data security and custom automation strategies (P01, P02, P03, P04, P05, P11, P13, P19). \add{For example, P01 and P05 led efforts to create guidelines for AI integration, aiming to reduce reliance on closed-source AI systems (P15).}  \edit{Participants} highlighted the quick AI development pace (P05, P16), \add{al}though some teams, limited by managerial gatekeeping, \add{had} only recently started experimenting \add{with} AI for brainstorming headlines (P16). \add{A tension was evident between disruptive innovation and the safeguarding of normative journalistic standards.} While some emphasized that newsrooms must remain ``innovative and modern'' (P16), others were skeptic\add{al} (P12, P13, P14), stating poor AI-generated designs or error-prone articles. Concerns ranged from AI illustrations considered a ``downgrade'' for newspapers (P12, P13) to criticisms of excessive AI hype~(P12, P14, P15). \add{P14 voiced that adding new tools to an already overwhelming workflow of applications is redundant if they offer no additional value. She emphasized the need to} focus on improving existing tools and processes rather than inventing something new (P14). Similarly, P12 assessed the hype and misplaced expectation surrounding AI: \textit{``They [journalists] think AI can do more than what it actually can!''}. Likewise, P15 critically observed that: 
\begin{quote}
Every program \add{now} has a button with AI, nobody asked for (...) They [tech actors] only want you to change your behavior to a point where you need it.
\end{quote}  
\add{Challenges were pronounced in local media}, where editors must combine both digital and print productions (P01, P05, P07, P08). P08 cited that\add{ \textit{``nobody is waiting to find a printed newspaper anymore''}}, \add{highlighting the constant shift to digital journalism.} \add{P05 emphasized the limited focus on collaboration among local outlets concerning the efficacy of AI integration in local practices. Participants speculated on tools to support collecting data and identifying newsworthy subjects online, \textit{``to see if there is something to write about or to put more effort into''}~(P20). This discourse suggested the potential for AI-supported innovation to augment local journalism data workflows.}

\subsection{What Are The Challenges Local Journalists Report When Working With Data and AI?~\textbf{(RQ2)}} 

Our analysis shows that journalists prioritize accuracy and reliability.  Participants highlighted \add{several} challenges, including a lack of skills \add{and} confidence, the risk of misinterpretation, the considerable effort required for analysis, as well as \add{limited} access and availability. Additionally, journalists reported little incentive to engage with data-demanding tasks, as \add{the work} is uncompensated (P02, P12), frustrating (P02, P05), and time-consuming (P02, P03, P05, P07, P10, P11, P12, P17).  \add{Unpacking the perspectives of journalists with limited expertise in data work helps us envision better future practices and design a more inclusive technology.} 

\subsubsection{Lack of Skills and Expertise} During the interviews, participants often expressed insecurity when asked to complete the online crime map investigation task. Several \add{respondents stated that} they would avoid such work due to limited training and a lack of data analysis skills (P03, P04, P05, P08, P12, P17). P05 emphasized \add{the difficulty of} not knowing how to even begin analyzing digital data. When discussing data expertise, many referenced ``Excel'' (P09, P16, P18, P21). For example, P16 described that \textit{``it was difficult''}, admitting she struggled with tasks \add{such as} sorting information and using commands in Excel. 

\subsubsection{Lack of Self-Confidence} We observed a lack of self-confidence when handling data, with participants often delegating tasks to technically skilled colleagues or data journalists (P01, P03, P05, P11, P17 P17, P19). P03 and P05 could not imagine evaluating raw web data themselves, and P03 said they would avoid reporting on data-heavy topics unless press offices provided exact figures.  P15 admitted to rarely working with raw data, instead asking others to explain \textit{``what does the statistics say?'',} relying on ``\textit{experts doing the heavy lifting on data}''. \add{In general,} several participants \add{regarded} their role as critically evaluating refined reports and shaping narratives rather than processing raw digital data. (P03, P12, P18) \add{They were not confident they would draw the right conclusions}. To illustrate, P17 described:
\begin{quote}
I believe they [colleagues, 20-30 years younger] could do it [data task] better than I do. I would not even trust myself to do that. I do not have the time to spend practicing a new technology. I am not patient enough, and I have no trust in these [AI] results. I would think it is not worth trying to find that data [online crime statistics], so I would not do it, or I would try to find someone else who can do it. With the tools I have, I would not try!
\end{quote}

\subsubsection{Data Volume} Working with large datasets (P02, P03, P05, P11) posed significant challenges. Participants noted the difficulty of sorting through vast amounts of information to identify meaningful insights (P11) and the high effort and extensive work required to do data journalism (P05, P15). Several highlighted the inevitable risk of human errors (P05, P17) due to fatigue, as longer processing times reduce concentration (P20). This raised concerns about inaccurate statistics, since the likelihood of overlooking important patterns increases with dataset size (P20). Journalists also face issues with inconsistent data formats (P02, P09, P13, P18, P21), receiving large volumes of documents from public entities that require conversion into analyzable content (P02, P09, P18). P18, though enthusiastic about the potential of data, doubted he would ever work with \textit{“big”} data. He explained that: 

\begin{quote}
When it comes to pure data processing, the problem is: how do I get the data? How is it available? How can I manage \add{an extensive} dataset to flag anomalies automatically? With the ``\textit{CumEx}'' cases [an investigation by several European news media outlets into a tax fraud scheme discovered in 2017], from the ``Süddeutsche Zeitung'', they \add{put in massive} data sets to analyze cash flows and understand the system. \textbf{As a local journalist, I will likely never work with a dataset like that}. 
\end{quote}

\subsubsection{Data Complexity} Since many journalists lack data expertise, a key challenge lies in translating numbers into clear, meaningful stories. Interpreting data correctly (P06, P14, P15), identifying patterns (P18), and  \textit{``extracting''} the valid \textit{``story behind raw data''} (P14, P15, P18, P20) is not straightforward. Some participants noted challenges in understanding statistics, as the same figures can carry different meanings across datasets that must be distinguished (e.g., accident reports ``appear in the same form in police statistics''~(P06)) (P04, P06, P13, P14).  P06 stressed the importance of knowing \textit{“what numbers actually tell”} and recognizing their \textit{``limits''}. He explained that: 
\begin{quote}
\add{You often assume that a number is something concrete, and its meaning must therefore be concrete. However, that is not necessarily the case. }
\end{quote}
P15 highlighted that \textbf{\textit{journalists are not statisticians}} and recalled that \textit{``people [journalists] do not fully understand data''}. Journalists alerted risks such as false correlations (P15), inaccurate results (P18), or misleading interpretations (P14) when ``unexperienced people mess with data they do not understand''(P15). In addition, P03 and P15 both cautioned that AI may oversimplify analysis, overlooking insights from the local context (P14). Finally, P14 said that ``linking data to establish meta\add{-}level relationships'' among sources was seen as another hurdle for non-experts. As P18 put it, journalists must build \textit{``a story with \add{a} valid data basis''} or risk writing \textit{``complete nonsense''}, highlighting the need for critical thinking, sensemaking, and strong data literacy to interpret statistics responsibly (P14, P15, and P18). 

\subsubsection{Data Access, Availability, Reliability}  When asked how journalists gather data, participants mostly cited public reports, governmental sources, institutional databases, or official press releases (P01, P02, P03, P04, P08, P09, P11, P13, P17). Local reporters rarely deal with open data or web scraping (P18). Many noted that data is not always public or complete (P02, P04, P13, P18), and \add{journalists often} waited for public releases or had to \add{contact sources directly} (P01, P03, P04, P13, P17, P18). Participants highlighted the lack of accessible online sources. P02 expressed frustration with delayed election statistics from officials, saying \add{local} newsrooms often receive reports at the last minute with little time for verification. \add{In general, governmental sources do not disclose full details online.} \add{Participants} reasoned that if Google cannot surface the data, automation likely cannot either (P02, P07, P18), since AI relies on public\add{ly} available source\add{s} (P08). 

In a different light, participants often noted that public sources use their own metrics \add{for assessing} the importance of \add{the} data they publish, and local journalists may interpret this information differently depending on the local context (P03, P06, P14, and P17).  In addition, we observed the importance and complexity of journalists' relationships \add{with} ``human'' sources (P03, P06, P08, P09, P12, P21), who provide supporting data (P06) and access to sensitive information (P18).

\subsubsection{Data Veracity}
As outlined earlier, limited access, availability, or lack of skills often forces journalists to rely on third-party reporting. This dependence on external sources raises concerns about reliability and transparency, as external data may be incomplete  (P04, P13, P21), outdated (P05), or inaccurate (P01, P13), leading to false conclusions and undermining journalists’ \textbf{trust} (P15). \add{P}articipants \add{critically assessed} the need to verify police reports, as police can make mistakes or evaluate things differently (P13, P14). Ironically, many journalists prefer waiting for official annual statistics (P01, P03, P04, P05, P09, P12), especially in crime reporting, because police are seen as having professional expertise (P03, P17). Yet, P13 and P14 noted that yearly police \add{records} may contain inaccuracies, hindering reporting quality. P13 further highlighted the legal risks of using external data, such as liability for inaccuracies and accountability in reporting: 
\begin{quote}
You \textbf{can not even trust the police}; not that they are lying to you. \textbf{They are just humans}, not so concentrated during work as I am, because my name is in the article. All the information I publish \add{links} to my name and my credibility.
\end{quote}

\subsubsection{Unreliable AI} Concerns about \textbf{trust, reliability, and bias} remain central. \add{Participants} stressed their responsibility for accuracy, accountability, and the pursuit of truth. They warned that AI’s errors and hallucinations (P02, P03, P04, P05, P07, P12, P15, P17, P18, P20) could undermine their credibility and the audience's trust when used without safeguards or quality control~(P08, P05, P13, P19). Participants highlighted the lack of control over AI’s decision-making (P03, P07, P08, P09, P13, P17) and its inability to capture local context (P01, P03, P05, P06, P12, P14, P17), which risks reinforcing stereotypes or even promoting racism (P13, P15, P17), particularly in sensitive reporting such as crime. Some feared biased systems could frame migrants or other groups unfairly (P13, P15, P17), effectively weaponizing narratives \add{that} can manipulate people's thinking (P15). Participants also highlighted the risk of overreliance on AI outputs (P02, P03, P07, P18, P21) because of its “\textit{black box}”~(P19) nature, warning that AI can overlook facts (P07, P17). AI’s false confident answers were also problematic (P06, P12, P18). \edit{Journalists openly criticized AI's lack of transparency (P16, P18), central to journalistic rigor.} \add{Journalists are held accountable for their mistakes (P08, P13).} Unlike journalists, who follow an ethical \textit{``codex''} (P13), AI lacks moral judgment (P13), failing to meet journalistic standards (P11, P12, P18). 

Journalists also raised concerns about an \textbf{efficiency trap} (P15) where verifying AI output \add{paradoxically} doubles the amount of work~\cite{ironies_of_automation1983} rather than saving time~(P02, P03, P05, P08, P15). P06 argued about the cognitive load when prompting to ``describe very precisely'' what AI should do. Participants \add{emphasized the importance of} prompt sensitivity when instructing an AI assistant, and \add{the need to learn how} to write \add{effective} prompts (P02, P06, P07, P11, P15, P21). \add{They also emphasized the value of} acquiring ``AI skills'' to use \add{AI} effectively (P05, P06, P11, P12, P13, P19). Some participants (P07, P12) found the reasoning processes of AI helpful, whereas P03 cited: 
\begin{quote}
I would not really need these individual steps [explanation of what AI is doing]. I would trust the machine. It is like with a laptop. \textbf{I do not know how it works, but it works}.
\end{quote} 
Despite skepticism, many participants acknowledged AI’s potential (P01, P02, P06, P08, P12, P18, P20), especially in data processing, and expressed interest in \textbf{explainable AI} (P07, P12, P18). Ultimately, journalists viewed AI as a supportive tool rather than a replacement (P01, P02, P06, P12, P16, and P19).

\subsubsection{External Threats to AI and News} Participants (P06, P12, P14, P15, P18, P19, P20, P21)  discussed how AI and automated reporting threaten journalism’s core business model, where users traditionally ``pay for entertainment and information''. \add{As consumption patterns evolve  \cite{reutersDigitalReport}}, automating news processes raises questions about the future of journalism (P06) and the benefits of the \add{audience}~(P06, P15). Journalists recalled that \add{readers} can \add{obtain} summaries directly on Google without visiting news sites (P06, P14, P18, P20), \add{which reduces readers'} willingness to pay (P06, P20). AI systems bypass paywalls, undermining revenue, with participants expressing concerns that \textit{``AI is partly taking away the media's livelihood''} (P14), and \add{worried} about declining business models (P06, P08, P15). P21 clarified that AI does not bypass paywalls but instead aggregates publicly available, often inaccurate, information, thereby amplifying misinformation. P15 doubted whether users even want AI-generated summaries, noting they were \textit{``implemented without asking the users''} and \textit{``it is hard to turn this [feature] off}''. As a mitigation strategy, journalists advocated for stricter polices (P14, P15) ``against (...) tech companies and data giants (P14)''. Nevertheless, P15 admitted that regulations are unlikely due to ``pressure from the AI industry'' and political factors. P20 \add{proposed} a business model \add{in which} newsrooms collaborate with AI providers to deliver reliable information in exchange for compensation. Despite uncertainty, participants emphasized that newspapers must preserve a strong \textit{``trust factor''} (P12, P18, P20).

\subsection{AI-Supported Reporting In Local Journalism: Participants Imagined AI Futures When Working With Data~\textbf{(RQ3-a)}} 

In this section, we focus on how \edit{AI-supported} reporting can \edit{augment} local news routines, including saving time when processing data, automating editorial processes, discovering news, fact-checking\add{,} and providing multiple perspective\add{s} for a story. Surprisingly, journalists \add{reported} that their work-related topics change frequently, \add{which restricts} the adoption of automation (P06, P07, P09, P12, P13, P19). \add{Therefore, we derive themes from participants' imagined scenarios when thinking about AI and data. Prior work suggests that AI systems are opaque and complex socio-technical mechanisms that untrained journalists struggle to untangle~\cite{JonesJonesLuger2022}. However, our participants \add{expressed} optimism regarding technical capabilities, expecting AI technologies to improve, and reinforced that the current state represents their worst performance. They viewed AI-supported reporting as a useful way to ``easily search websites (P18)''  and identify ``exciting (P20)'' story directions}.

\subsubsection{Personal (Time Saving) Assistant} All participants considered the automation tools as an \textit{``added advantage''} to their current workflows, saving time on routine and undesirable work (P01, P08, P09, P10, P11) and enabling quick journalism to collect and produce news fast (P08, P09) as journalists might \textit{``never have time for a story''}~(P05, P06). Participants considered AI \textbf{a personal assistant} (P20) to share the workload, prioritize and schedule tasks, \add{and auto-reply} to emails (P12), leaving more creative time to write complex stories (P05) and \add{conduct} in-depth reporting (P05, P12). Similarly, P01 tagged AI as a \textbf{little new colleague}, stating that: 
\begin{quote}
I can give all the work that nobody else wanted to do (...), and you can say, ``Find this, or create and visualize it for me. (...) Just do it''!
\end{quote}

\subsubsection{Automated Editorial Processes} Another important observation is the desire of journalists to offload some editorial processes to automation. Some participants even discussed\textbf{ a fully automated editorial system} (P01, P03, P05, P12, P15, and P18), which includes \textbf{creating newsletters} using keywords, customized \textbf{article rewriting}, \textbf{embedding graphics and social media posts}, and designing \textbf{custom newspaper layouts }(P18). AI can support \textbf{editorial decisions}, such as determining which stories should appear on the front page of a newspaper (P02) or updating the meta-descriptions of high-performing articles online (SEO) (P07, P14). Moreover, participants considered using AI to manage social media (P02, P14) and online communities (P01, P04), as well as to edit short videos, create reels, and generate headline hooks for articles (P04). P01 described the process as cumbersome:
\begin{quote}
Currently, we still \add{perform} all of these tasks manually. \add{An email arrives with the press release; [we] reword it, and [we] look for pictures.} In my opinion, \textbf{that is not really journalism}!
\end{quote}

\subsubsection{Monitoring \& Discovering News} Another important them\add{e} is the continuous need of journalists to monitor and discover news and ideas from multiple sources (P10, P14, P20). AI-supported reporting can \edit{increase newsrooms' competitiveness capacity} by quickly bringing compelling stories to life (P01, P02, P03, P04, P10, P20). Participants explained how AI-supported reporting can assist in analyzing various unstructured content scattered online (P18) and computing statistics (P01, P02, P03, P07, P08, P10, P15, P18, P20) by autonomously navigating the web (P02, P08, P18). In addition, participants imagined to use the tools to: \textbf{scan} social media accounts for ``\textit{cool trends} (P01, P02, P04, P06, P07)'',  periodically \textbf{check, compare} and \textbf{aggregate content} from public sources alerting the newsroom about regional trending topics (P01, P02, P06, P07, P09, P10); \textbf{collect} latest published articles from competing news outlets (P01, P02); or go through own newsroom's archives and analyze occurrences of previous reports (P02, P09, P13, P14). According to P02, P06, and P15, the ability to visualize data instantly, quickly prototype and test ideas can help verify ``\textit{hypothesis} (P02)'' instantly and assess whether there is a story behind the data. P20 emphasized the simplicity \add{of collecting} the information and \add{updating} citizens (e.g., identifying risky areas, ``accidents'' hot-spots (P06)), ``\textit{without writing a single line of code} (P02)''. Overall, participants believe AI can reduce researching time (P01, P04, P07, P08, P11, P12, P13) and help find sources on a hyper-\textit{``local level''} (P01, P03). 

\subsubsection{Automated Fact Checking} Verifying facts is one of the most important journalistic tasks~(P01, P03, P05, P13, P17). \edit{Participants emphasized the heavy burden of fact-checking (P04, P05, P07, P16, P18), since comprehensive analysis was both a major challenge and a necessary safeguard against inaccuracies (P09).} We observed that journalists do not trust AI for fact-checking~(P03, P12, P17); however, they identified its potential to detect fake information (P18, P21). Most participants considered using AI to compare sources (P07, P09, P14, P19) and to verify their trustworthiness (P01, P07, P19). As an investigative example, P07 explained how to combine press releases, public talks, and social media posts of researchers to mine their opinions and finally link their funding sources to private companies. Overall, participants highlighted that using AI feels like \textit{``having a million pairs of eyes to watch over data''} (P01, P05).\\

\subsubsection{Interactive Data Processing}~AI-supported reporting tools transform unstructured data (P07) into meaningful in-depth and clear analyses (P04, P07, P12, P13, P14). In this context, P03 mentioned that these tools would offer a completely revamped approach to accessing and embedding data in reporting. Inspired by the crime investigation task, journalists suggested that AI can quickly classify online reports, link cases, evaluate statistics, \add{and} compare figures (P02, P07, P08, P10), enabling journalists to \add{gain} insights \add{into} correlations  (P14) and causality of events (P01, P14). AI-\add{supported tools} empower local newsrooms to create stories never thought possible or previously unfeasible due to \add{a} lack of resources or time constraints (P02, P06, P08, P09) and help them ``\textit{\add{not to} be afraid to use data anymore} (P21)''. However, participants stress the importance of human oversight.

\textbf{Visuals Reporting:} Participants believed that AI tools can assist and enrich their stories quickly with easily graspable graphics, especially as AI can remove convoluted math or language phrases (P01, P12, P06, P09). 

\subsubsection{Angle Reporting: Shift Story's Perspective} AI-supported reporting introduces several angles to storytelling, encouraging journalists to align their writing with readers' interests (P02, P03, P05, P15, P18). These tools can help analyze users' reading history (P15) and bring the audience closer to journalists (P02, P05). \add{Participants hold} different viewpoints, with some believing that AI \add{will} bring a ``\textit{novel approach to storytelling}'' (P03) and sharpen the authors' perspective on a story (P16). \add{In contrast,} others \add{remain} hesitant, especially considering that AI cannot \add{fully} understand local context (P06, P17). There is a ``discrepancy (P17)'' between journalists' experience on what they find interesting to report, and readers' actual interests (P06, P15, P17). P17 explained that:  
\begin{quote}
I am not convinced that my reading [the view of the word] is really correct, so I would not trust \add{myself}, and I would not trust the AI as well, even if I am the one who decides how the AI is supposed to read the news.
\end{quote}

\subsection{AI-Supported Reporting In Local Journalism: Socio-Technical Opportunities (Redefining Local News Routines) ~\textbf{(RQ3-b)}} 

Our conversations reveal that journalists would seize the \edit{AI-supported opportunities} to undertake additional complex tasks (P01, P05) with a focus on: (i) character-driven storytelling ; (ii) in-depth journalistic research; (iii) investigative and fact-checking work. P08 explained that ``\textit{media is only [now] starting to understand the revolution AI would bring}''.  As AI is faster and more accurate than humans (P05, P08), journalists should `\textbf{`redefine their role in modern-day journalism}''(P08).  P08 explained that:   
\begin{quote}
By letting AI do some of our tasks, we could use the free time to concentrate on what we do better [than AI] (...) research, talk to people, and ask sources.
\end{quote}
Several participants reflected on how AI-supported reporting might impact journalists' self-perceived roles, considering both opportunities and challenges as AI tools do more data work (P01, P03, P04, P06, P08, P10, P12, P17, P19, P21). 

\subsubsection{Focus on What They Do Better: Human-Contact Reporting} Journalists reported that they would devote more time to \textbf{character-driven reporting} \textbf{\add{that focuses on} human connections} (P01, P04,  P05, P08, P09, P12, P21). Some participants considered the ability to talk to people \textit{``as a key strength of journalism (P01, P08, P21)'' }and a \textit{``unique selling point in the future (P04)''}. Accordingly, they argued that the audience cherish\add{es} ``human stories'' (P08, P10), which involve direct human contact (P09) and ``\textit{emotional reports}'' (P16). Readers not only value the story but also the writing style and the unique perspective of the authors (P19). They appreciate articles written by humans instead of AI (P12). Additionally, local journalists feel a responsibility to conduct research (P09), contact sources (P16), \add{conduct} interviews (P12), maintain \add{relationships} with readers (P19), and report stories in a local context (P06, P17). P08 framed that as a ``\textit{classical role}'' of a reporter: 
\begin{quote}
\add{One of our reporters entered Israel. He is talking to people on the streets, asking how they are feeling about the war, what they have experienced in the past days, and where this war is going.}
\end{quote}
 
\subsubsection{Local In-Depth Research Reporting} Participants considered using AI tools to \textbf{conduct more research} on the local level and \textbf{dive deep into stories}. In general, journalists need to pull stories from the data (P01, P02, P06, P07, P10, P14, P15, P18, P21) through the lens of local context (P06, P14, P16) and make the articles ``interesting for someone to read'' (P01). Well-researched stories perform better than simple articles (P01, P12). However, conducting research is a time-demanding task (P04, P05, P12, P19, P16, P07). Automation can alleviate the time pressure, enabling journalists to support readers with in-depth local information (P01, P03). P12 stressed the added value of  data, enriching (P03, P16) and making the articles more \add{appealing to} readers. He criticized several local media outlets \add{for publishing articles without providing} details on statistics. Statistics offer insights, explain certain phenomena to readers, assist them in making more informed decisions (P15, P01, P12), and even spark public debates that push policymakers to consider change (P01). P01 emphasized the role of official reported statistics in local news in raising awareness against women's violence. As newspapers struggle to retain readership (P01, P02, P06), AI-supported reporting can ``position the newsroom ahead'' of its competitors (P19). As P19 highlighted, if journalists manage to create data-driven stories in the right way, they will ``give readers something other newspapers do not [provide]''.

\subsubsection{Towards Investigative Reporting}  Participants mentioned the importance of data and  journalism's profound role to shed light on wrongdoings (P12), to spot anomalies within a city (P14, P18), to promote social values (P09), and to contribute to an informed society (P01, P03, P06, P10, P15, P18). Journalists feel responsible to address under-reported areas (P08), uncover ``scandals'' (P12), or \textit{``unearth information not yet publicly \add{known}''} (P08). Participants stressed the grounding role of humans in publishing investigative stories (P12) and reporting on cases happening behind closed doors (P08), since ``not everyone gives sensitive information voluntarily''~(P18) and \textit{``AI cannot go into a room full of corrupted politicians}'' (P08, P12).

Furthermore, considering the quick spread of misinformation as a result of the misuse of generative AI (P01, P04, P05, P18, P21), journalists emphasized the need to dedicate more time to \edit{verifying AI generated content} (P01, P04, P05, P08, P15, P18, P20, P21) and monitoring \textit{``what is AI doing?} (P05)'', alluding to the significant role of  human agency in automation. P01 expressed the concern that the \textit{``internet might soon become obsolete when it comes to information and news, as no one will know \add{whom} they can believe anymore''} and speculated on having a team of  journalists dedicated to checking the local regions, consulting their sources to gather accurate intel (P01, P08), and ensuring the integrity and reliability of reported news. On this basis, the local newsrooms should position themselves as \textit{`reputable''} trustworthy sources of information (P12, P20). To conclude, P15 illustrated a hypothetical scenario:  
   
\begin{quote}
In a suburb, heavy with migrants, a far-right political party wins the majority of votes. The AI tools might flag the ``suspicious'' correlations among voters. However, simply looking at the data cannot explain why the people who would suffer the most would vote \add{in} that \add{way}. You can retrieve the data easily. However, you always have to [investigate] and ``ask experts to give proper explanations''. 
\end{quote}

\subsubsection{Journalists' Autonomy and Agency}

\add{
Participants expressed concerns about losing autonomy over automation and the inability to verify the AI-generated outputs (P03, P04, P07, P11, P16, P21). They emphasized the need to ensure that AI remains under human control~(P05, P09, P21). This sense of agency correlates with their perception of doing meaningful work~\cite{MollerOneSizeFitsSomeJournalisticRole2025}. Over half of the participants acknowledge the potential risks of job displacement, fearing that \textit{``AI will destroy jobs''}~(P11). Journalists stressed the need to recalibrate their tasks to maintain autonomy. Most respondents perceived their own roles as less at risk from AI compared to their colleagues doing conventional desk-bound work. However, when these concerns are filtered through the ethical and accuracy issues associated with AI-supported reporting, P12 substantiates that \textit{``If you want to have a good [journalistic] product, you still need humans in the long term''.} P17 shared similar thoughts, stressing the complexity of journalism and its non-deterministic nature: 
\begin{quote}
AI cannot replace our work at this time. I do not know what will happen in five years, but at the moment, I want to see more respect for our work. It is hard to get that information, and not everyone realizes that you do not just need to ask [plain] questions on your phone and instantly get the answers. [Journalism] is not that easy! 
\end{quote}   
Audience trust in human journalists is a key advantage newspapers have over AI journalists (P12). Preserving autonomy, agency, and independence in reporting is important to participants' concept of journalism and their role.
}

\subsubsection{Bridging Socio-Technical Knowledge Gaps} 

Participants expressed concern that not everyone can keep up with newsroom innovation, especially older colleagues who are less comfortable with technology (P01, P02, P12). Participants highlighted the generational gap and \add{emphasized} the need \add{for} adequate support \add{in} adapting to AI tools (P15, P16). Many reinforced the call for more training and hands-on experience with AI (P06, P08, P12, P13, P16, P19). Some participants found newsroom training helpful (P12, P13), others found it insufficient (P11, P19). \add{P15 described} AI-supported reporting as an entry point into data journalism and a risk if untrained journalists misuse data. Journalists acknowledged the need to develop data journalism (literacy) as an essential future competency (P11, P15, P18). They signaled a need to sharpen critical thinking (P01, P03, P06, P14, P15, P18) and to refresh their outdated data skills (P16). P06 concluded the status quo of \add{AI in} local news and the challenges of future journalism: 
\begin{quote}
(...) a question of what the future holds for me professionally, or what the future holds for everyone working in journalism. Will there be further dissemination and technical progress through AI systems? Ultimately, when it comes to work, AI should be \add{viewed} as a tool, which it certainly is, and in that respect, we must \add{strive to stay up} to date and use these [technologies] in a profitable and beneficial way.
\end{quote}

\section{Discussion}


Our research reveals that most local journalists do not have substantial expertise in working with large volumes of ``raw'' digital data, and many perceive this task as beyond their scope\add{~\textbf{(RQ1)}}. Besides limited expertise, insufficient time, and the complexity of data-related work, the study identifies a pervasive lack of confidence among journalists in their ability to perform data-intensive tasks. We \edit{highlight} the challenges~\textbf{(RQ2)} and self-perceived opportunities arising from AI data support \add{in} local journalism, \add{explicitly} focusing on \add{AI's envisioned future} benefits~\textbf{(RQ3-a)}, the added value to local news, and its socio-technical impact on the role of journalists in informing the public with reliable content~\textbf{(RQ3-b)}. Our findings \edit{suggest} that AI has not yet drastically transformed established journalistic practices, and most participants do not believe that AI-supported reporting will profoundly change how journalists work. However, it may reduce repetitive and time-consuming tasks, allowing journalists to focus on what they do best: reporting that involves human contact, verifying facts, thoroughly investigating topics, and providing readers with valuable, statistically substantiated narratives (see \cite{thaslerKordonouri2025automatedUK}). 
\add{The relationship between journalism and AI, their societal impact, journalists' values, and their audience is in a constant state of mutual reshaping~\cite{Lewis2025GenAIDisruptiveJournalism}. This process involves redefining journalists' roles, reframing journalists’ expectations towards technology, and the consequent disruption of traditional professional journalistic norms and identities~\cite{Lewis2025GenAIDisruptiveJournalism, guzman2024what}.}

In the following \add{sections}, we first corroborate our findings with prior works and then address the limited awareness of data in local journalism in Germany. However, we believe our findings are more broadly applicable. Therefore, we propose that future research investigate\add{s} how to cultivate data-thinking competencies among local journalists. Finally, we reflect on the broader vision of AI-supported reporting for local \add{media} and propose technical recommendations to assist local journalists with their data endeavors.

\subsection{Automating News Locally: Do-It-Yourself}

In line with prior work~\cite{DiakopoulosAPReport2024, Cools2024PerilsPerceptions, AspenDigital_AI_GrowingRole2025, DhaeseleerAIDivides2025, MollerOneSizeFitsSomeJournalisticRole2025, Simon2024AIinNews}, we find that local journalists already had an initial point of contact with AI, primarily involving text-based editing, sensemaking, and simple research. Our results show that they rarely employ AI for extensive data-related \add{tasks} such as online data collection, news discovery, or content verification due to \add{concerns about} trust in AI, data safety, and \add{the potential for} hallucinations~\cite{kalai2025HalucinationsLanguage}. This observation aligns with the existing literature on AI and journalism \cite{Cools2024PerilsPerceptions, DhaeseleerAIDivides2025, CoolsFromAutomationToTransformation}. Interestingly, local journalists expressed strong interest in AI's potential to support their reporting by supplying data, aggregating trending sources and online content, and identifying \textit{``newsworthy''} stories computationally by extracting patterns from data. However, while some participants could envision using AI to process online data, most of them never leveraged AI for this purpose. \add{This task was not part of their usual newsroom roles}. This finding suggests that local journalists possess limited awareness of the full scope of AI capabilities.

\subsubsection{Do We Really Need AI-Supported Reporting?}

After exploring our prototypes, journalists praised the simplicity of AI-supported reporting for \textbf{collecting digital data on their own}, without requiring \edit{deep} technical expertise, an activity previously considered impossible or not feasible due to time or resource constraints. The scarcity of AI data use cases in local journalism suggests that local newsrooms are still in their early \add{stages} of experimentation, and local reporters have not yet \add{extensively} engaged with such approaches. Similarly, prior work documents experimental levels of AI in the newsroom~\cite{DhaeseleerAIDivides2025, EderFallingBehindLocalJournalismStruggle}. To this end, local newsrooms must consider how to better incorporate and embrace AI in their processes by rethinking how they produce news \cite{Kielland_2023_NewsCarousel}. The influence of large technology companies on news organizations \add{and their future infrastructure dependency~\cite{sjobvaagBeyondPlatforms2025capture}} will continue to grow~\cite{Kielland_2023_NewsCarousel, SimonUneasyBedfellows2025, Simon2024EscapeMeIfYouCan} unless newsrooms establish internal innovation labs\add{. Critics argue that newsrooms} need to develop practical AI applications and tools~\cite{EderFallingBehindLocalJournalismStruggle} to sustain their growth and \add{independent~\cite{MollerOneSizeFitsSomeJournalisticRole2025}} flow of information, which supports democratic societies. However, the need for automation in local journalism warrants closer examination.

Ultimately, Stray reminds us, integrating AI depends on the opportunity cost of automation tools relative to existing reporting methods and the core role of journalism~\cite{Stray2019}. Asking the source directly could be more advantageous than \add{relying on} automated data processing~\cite{Cohen2011_ComputationalJournalism}, especially when multiple perspectives emerge on what is considered right or wrong due to numerous extrinsic\add{,} complex factors that can influence decisions, such as the political climate or contextualized decision of policymakers, which are impervious to AI \cite{Stray2019}. In this case, the role of a local journalist is prominent in ensuring that information is conveyed \add{accurately} to the audience and does not mislead readers, especially when dealing with sensitive topics.

Nevertheless, in this paper, we report that AI-supported tools are perceived as providing added value to journalistic data workflows, allowing local journalists to collect, structure, and verify \add{datasets} autonomously, \add{thereby} enabling them to engage in more data-driven stories. These results indicate that local journalists without technical knowledge can use AI-supported reporting tools to \textbf{quickly interpret complex data}, thereby \textbf{increasing the speed} and \textbf{expanding the scale} of their reporting~\cite{VeerbeekFightingFireWithFire, besjonHendrikCHILBW2025}. This process, in turn, \textbf{saves time} and alleviates publisher pressure~\cite{thaslerKordonouri2025automatedUK}, promoting a healthier local news ecosystem. \add{Moreover, we corroborate prior studies indicating that journalists could utilize AI to reduce repetitive work and redirect their attention to in-depth investigative journalism~\cite{Moller2024, schapals2020assistance}}. We highlight \add{further} insights into the challenges local journalists \add{encounter} with data. \add{These findings} can inform future tools to support local media\add{'s} data processes and help them \add{produce} higher-quality stories.

\add{Furthermore, our analysis raised ethical questions. As AI advances, debates over which tasks journalists can delegate to AI and the level of oversight they should retain intensify. For every journalistic task automated, the morality of such actions is partially transferred to the algorithms~\cite{DoerrHollnbuchner2017, SinaKordonouriJournalisticAgencyAI2025}, thereby redefining journalists’ autonomy~\cite{cools2023AutonomyAgencyLevels}. Distrust in AI tools due to unreliability, limited transparency, and a lack of control over automation prevents journalists from using AI to work with data. Journalists see it as an added risk to core journalistic standards informed by their non-technical background and limited experience with digital data. Their emphasis on control and transparency highlights the need for human-in-the-loop approaches to strike a balance between AI utility and ethical integrity.}

\subsubsection{Trust in Local News: The Bigger Picture}

Study participants argued that integrating AI into reporting would enable them to work with substantially larger datasets. They expected AI-supported reporting to enrich the newspaper content with evidence-based insights, \add{thereby supporting} their readers \add{and} strengthening trust in the news. In this context, data can play an instrumental role in shaping social issues~\cite{JarkeReport19}. When reported correctly to the right audience, data-driven stories can influence \add{their} decision-making~\cite{ShullerDataLiteracyFramework2019} and sometimes \add{lead to} changes \add{in} established laws and policies.

Given the limited research on data-related local journalism, we explore local journalism’s ambivalence toward digital data from a socio-technical perspective. Our findings corroborate previous works analyzing AI usage patterns among journalists across the Netherlands~\cite{DhaeseleerAIDivides2025, Cools2024PerilsPerceptions}, Denmark~\cite{Cools2024PerilsPerceptions, CoolsFromAutomationToTransformation}, and Belgium~\cite{DhaeseleerAIDivides2025}, \add{which are} located near Germany and share similar news ecosystems. Although \highlight{these prior works} did not focus on local journalism or non-technical news workers, D’haeseleer et al. reveal that, in general, AI is not disrupting journalistic workflows~\cite{DhaeseleerAIDivides2025}. \citet{DhaeseleerAIDivides2025} report that most journalists \add{primarily} use AI tools for text support, \add{a finding} also noted by \cite{Cools2024PerilsPerceptions, CoolsFromAutomationToTransformation} and other previous surveys~\cite{DiakopoulosAPReport2024, beckettYaseen2023generating}. In addition, all studies show that several journalists contemplate using AI for information gathering; however, only a limited number of journalists consider using AI to verify content. In our study, we also found \add{that} a \add{significant proportion} of participants weighted the use of AI-supported tools for finding and exploring sources.

Contrary to earlier reports, our research indicates that journalists plan to use AI for factual verification to validate leads and sources, considering that future advancements in AI \add{may} reduce hallucinations, improve accuracy, and enable more \add{effective} source control. In doing so, local journalists aim to combat misinformation and fake news, establish\add{ing} their newsrooms as reliable and credible sources of information. Their motivation \add{suggests} that local journalists \add{may} play a prominent role as fact-checkers, as several fake local media outlets are increasingly using AI to generate \add{false information}~\cite{Fischer2024_DarkMoneyNewsOutlets}. Journalists must therefore acquire the necessary skills (i.e., data literacy~\cite{ShullerDataLiteracyFramework2019}, computational literacy~\cite{Wing2006}, and AI literacy~\cite{ShullerDataLiteracy2022, DhaeseleerAIDivides2025, CoolsFromAutomationToTransformation}) and build toolkits to manage these data-intensive tasks. However, in \edit{local} newsrooms with limited resources, this goal remains elusive unless supported by external sources~\cite{EderFallingBehindLocalJournalismStruggle}, \add{such as} researchers or the open-source community. Although prior research has explored how to create tools in collaboration with journalists~\cite{tseng2025ownership}, it often neglects the implications for local news. These findings highlight opportunities for HCI research to explore how to develop reliable tools \add{for} verify{ing} facts and detect\add{ing} claims \add{among} local journalists through participatory design~\cite{bodker2022participatory} and co-creative approaches. These strategies could translate local journalists' needs into functional applications.

Interestingly, our findings suggest that most local journalists trusted third-party reports and local authorities (as data at \add{the} local level is sparse~\cite{StalphDataLocalJournalism}), such as police statistics and government reports\add{. However,} some mentioned the inaccuracies in these reports. The police were a reliable source of information in the past. However, the German Association of Journalists recently urged reporters to do their own research \cite{DJV_2024_RechercheUnverzichtbar,Deutschlandfunk_2019_PolizeiPrivilegierteQuelle,Netzpolitik_2022_Polizei_PrivilegierteQuelle,Uebermedien_2022_Polizei_UnzuverlaessigeQuelle}. AI-\add{supported} methods enable local journalists to process data independently and draw their own conclusions from statistical comparisons. To support journalists' claims, AI-supported tools could provide additional sources, enabling journalists to verify content and improve the overall trust in the news publisher. HCI research can further identify optimal interaction models that place users' priorities at their core, facilitating transparent decision-making in synergy with journalistic values.
\subsection{I Am Not a Data Person: Data Literate or Data Reliant}

Our results indicate that journalists' challenges when working with data may influence their interaction with AI. Journalists who are not equipped with data literacy~\cite{ShullerDataLiteracyFramework2019}, computational thinking~\cite{Wing2006, diakopoulos2024data}, or \add{who} have limited statistical knowledge \add{may} not fully realize AI-supported reporting benefits. Unless journalists are proficient in interpreting the meaning behind data or evaluating the validity of automated output, automation and AI \edit{may} not have a substantial impact on their work. \add{We do not suggest that all local journalists become data experts capable of creating complex data pipelines. Instead, our goal is to advocate that journalists should at least understand how to search, query, and verify digital data, enabling them to cross-check information effectively. Journalists should learn how to ask the right questions when analyzing data and draw meaningful conclusions. They should know what is possible in theory, even if they cannot do it themselves. Above all, journalists should recognize the limitations of AI-supported methods. Hence, equipping local journalists with this necessary agency when exploring data strengthens the community watchdog reporting.}

Additionally, journalists can only translate part of the collected data into stories because data without \add{a} local news context has limited meaning and utility. \citet{JarkeReport19} suggests that using data in the proper context can unmask representations of social issues. However, if local journalists lack the skills to critically evaluate and assess data's veracity, they may spread misinformation and present a distorted reality. Such practices would not adhere to journalistic values and principles~\cite{Nishal2025Values}. Moreover, data stories might negatively impact the readers' comprehension of the news~\cite{TooManyNumbers2024}  if numbers are not properly curated. This evidence reinforces our \edit{claims about} the need for journalists to \add{possess} a statistical understanding to simplify numerical language for readers.

These observations raise the questions of \textbf{1.} How to promote data literacy (which could complement AI literacy) in local editorial teams? and \textbf{2.} How to design tools so journalists with limited data expertise can use them responsibly in the local news context?

Our analysis shows that data literacy can influence how journalists interact with AI, interpret automated analysis, and correctly convert data reports into local knowledge. To our surprise, some local journalists did not consider their work data-driven, even though they use data and statistics \add{in} various contexts, such as elections or health reports. These statements primarily derive from participants\add{'} self-conception\add{s} of the term ``data-driven'', their perceptions of their role\add{s} in the newsroom, \add{and the extent to which they are exposed to big data sources}. Our results suggest \add{that some} local \edit{reporters} have limited data literacy \add{or limited professional experience}, particularly when working with raw digital data. \add{These findings are consistent with those of earlier studies~\cite{ScottNumbersInNews2002, ScottNumericalLiteracy2003, NeillPassiveJournalists2008, BorgesRey2020}, suggesting that journalism undergoes only minor and gradual changes}. The limited data competency indicates a shortage of \add{relevant} education~\cite{Figl2017DataJournalismSmallNewsrooms}.

Unfortunately, there are no systematic studies examining the university majors of journalists~\cite{vos_craft_2017_country_report_us}. The 2016 Worlds of Journalism Study did show that journalists are generally highly educated, but it did not provide details on their fields of study. In the U.S., journalists with a college or university degree typically majored in journalism (59.2\%) or communication (15.0\%)~\cite{vos_craft_2017_country_report_us}. As a starting point for assessing the situation in Germany, we looked at the educational backgrounds of participants in the Henri Nannen School of Journalism's training program, one of Germany's most prestigious journalism schools, likened to an elite academy for investigative and feature journalism. To understand what leading journalists in Germany have studied, we analyzed the members of the school's 40th training cohort~\cite{nannen40lehrgang}, the most recent program for which \add{data were available}. Most group members had studied humanities and social sciences, including German studies, sociology, political science, history, philosophy, journalism, and media culture. None had a background in data science, computer science, mathematics, or related fields, and only one person had studied psychology, which likely involved advanced statistical training as part of their program. 

\add{Maier argues that journalism schools should include more data training in their programs to build confidence among young journalists~\cite{ScottNumericalLiteracy2003}. Maier links journalists' experiences with numbers to their self-perceived beliefs and capabilities in handling numerical data. Accordingly, journalists are convinced they are not \textit{``numbers people"} and see themselves as lacking the skills to work effectively with numbers, even though much of their professional work requires these particular competencies~\cite{ScottNumericalLiteracy2003}.}   

These findings indicate that journalists, media experts, and researchers should prioritize reinforcing data and computational literacy within local journalistic teams to benefit from AI-supported reporting. We reported that most journalists delegated the data-intensive tasks to a more technically experienced colleague. In this context, persuading local journalists to interact with digital data independently remains challenging. We attribute this data reluctance to the limited data training and insufficient experience~\cite{Appelgren2014DataJournalism}, which contribute to a lack of self-confidence when working with data, as we have documented.  \add{Future work can investigate how AI can support data training in a ``chat as learning'' framework~\cite{RanYu2025ChatAsLearning} where AI could guide and teach users relevant skills and data concepts. }

\subsection{Implications for HCI and Design Recommendations: Vibes Are Not Enough If You Lack Local News Context}

The central topic of discussion during \add{the} interviews was the lack of local context in AI systems, \add{particularly} when finding important patterns in a dataset. Combined with the inaccuracies and hallucinations encountered by local journalists, this erodes trust in such systems. Consequently, automation without clear safeguards will harm publishers, weaken news reliability, and undermine audience trust \cite{Graefe2016_GuideToAutomatedJournalism, WarrenShowTheWork2025}. Therefore, local journalists need reliable tools to explain their reasoning \add{behind} their \add{decision-making}. Most importantly, these tools must resonate with the local news context.

One key issue when designing AI tools for journalists is the necessity for an evaluation benchmark for news processes~\cite{li2025benchmarkingJournalismTasks, li2025ecologicallyvalidllmbenchmarks}. Validating journalistic tasks remains challenging because AI systems \add{struggle to} embed local news context accurately. As our participant reported, news reporting is sometimes subjective, and generative AI cannot capture the local news nuances. \add{E}ven if journalists can compress their complex tasks into simple prompts, AI could draw unintended conclusions. Journalistic tools praised for their summarization and research capabilities are overhyped, \add{fail to} provide valid assistance, and typically perform poorly with large amounts of content~\cite{Schellmann2025_AItoolsForJournalism} e.g. these tools miss out on important facts during transcription~\cite{CarelessWhisperHallucinationsHarm}. 

Guided by these shortcomings, we propose recommendations that HCI researchers and news organizations can use to develop tools for local journalists in accordance with their ethical principles. Our proposals balance automation with human oversight, which our participants identified as important. \add{These insights resonate with existing literature that advocates for a hybrid approach~\cite{diakopoulos2019automating, SinaKordonouriJournalisticAgencyAI2025}, in which AI and journalists collaborate to automate non-critical processes in accordance with ethical guidelines.} Furthermore, our findings show that local journalists seek solutions that \textbf{deliver real value}. \add{Participants highlighted an abundance of tools in their workflows. AI innovation introduced new processes without entirely replacing existing activities, increasing the complexity of digital journalism~\cite{Figl2017DataJournalismSmallNewsrooms}. Consequently, our informants} preferred simple tools \add{that} they can easily incorporate into their existing workflows (see also a related framework for designing human-centered tools for journalism~\cite{hagar2025tinytools}).

\subsubsection{Shared Prompt Configuration and Automation Support}

Journalists mentioned that older colleagues struggle with prompting and reported that they usually share the best prompts within their teams. This behavior indicates that some journalists require guidance in writing suitable prompts, \add{as the specificity of prompts can skew the expected results~\cite{ronanki2024requirements}. Current work shows an increasing trend toward making AI solutions more deterministic~\cite{he2025nondeterminism}}. A shared prompt library enables users to customize existing prompts or obtain template suggestions. This approach can improve automation efficacy and assist users in getting started. Newsrooms can integrate tools like Chainforge~\cite{arawjo2024chainforge}, which offer visual guidance to validate prompt accuracy and improve usability. Requirement-Oriented Prompt Engineering~\cite{MaEtAl2025_ROPE} is another paradigm in which AI systems can ask users to improve their prompts iteratively, \add{guiding} them to add details to complete the context. Additionally, in an AI-supported reporting system, journalists can share automation workflows and instructions within editorial teams and collaborate to create, combine, and reconfigure the automation scenarios while investigating a topic. This method could lower \add{the} initial barriers to automation for journalists with limited computing knowledge \add{and facilitate journalists' collaboration. Some participants viewed AI as a new avenue for collaborative engagement that can provide guidance or assistance. Collective approaches to utilizing AI systems for collaborative work benefit newsrooms~\cite{xiao2025genaiindividualusecollaborative}. With the new group chats released by OpenAI~\cite{openai2025groupchats} (multiple users interact on the same AI chat interface), understanding how these practices impact journalists' agency is important to determine the future directions of collaborative work in the newsroom.}

\subsubsection{Interactive Data Interpretation \& Verification}

Journalists reported that they struggle with mathematical or statistical concepts. Data visualization and interpretation, in the form of interactive graphics or tables, provide additional insights that explain the results and anomalies more \add{straightforwardly}. This approach makes it easier to find correlations because some local journalists may have difficulty using Excel, as documented. These functionalities could ease the cognitive load when processing and analyzing data, assist journalists in \add{their} research, and \add{provide} them \add{with} a better way to communicate data to the public. \add{Research has shown that journalists may be reluctant to reject AI output~\cite{nishal2024envisizoningapplicationsimplicationsgenerative}. These interactive approaches offer more transparency, prompting journalists to reevaluate the AI remarks.} 

Additionally, we reported that participants valued sources of the generated responses, allowing them to trace back to their data origin. An interesting future direction is to understand the needs and preferences of local journalists regarding AI reasoning traces \cite{chainOfThoughts2022}. Journalists often reported that the system invents false facts and is confident in the veracity of fake information~\cite{kalai2025HalucinationsLanguage}, even when instructed otherwise (P06, P12, P18). Therefore, \add{AI-supported} systems should provide insights into the uncertainty of their responses. \add{They could report} a hallucination score or a summary. \add{Such logical} traces could help journalists \add{assess the reliability of} the output and \add{add substantial} context to the statistics. Recent work~\cite{sun2025explainingsourcesuncertaintyautomated, Farquhar2024} explains insights on how to \add{derive} such scores through statistical evaluation or text entropy.

\subsubsection{Local-Context-Aware Angle Reporting}

Participants preferred using AI to adjust their stories to better align with readers' interests. A study shows that LLMs can offer alternative perspectives on press releases, exposing journalists to multiple viewpoints, as their own may be biased or incomplete \cite{PetridisEtAl2023_AngleKindling}. Our work also shows that journalists seek more control over the reporting angles, including local news context. Local journalists reported that some knowledge is not available online. To overcome these limitations, we invite future work to investigate the potential of integrating external knowledge bases as a knowledge graph (KG) into AI-supported systems for local contextualization. KGs include multiple relations between entities~\cite{Hogan2021KnowledgeGraphs} and can even validate or invalidate relationships as human knowledge continuously evolves~\cite{rasmussen2025zep}. Newsrooms can repurpose their previous reports or archives to construct a knowledge base (P02, P14), enabling local journalists to determine whether any news domain \add{remains} underreported or where the story's focus should be next. A customizable interface that incorporates external context could add nuance to local stories~\cite{NishalCDN25}. Such systems can provide a ``newsworthiness score''~\cite{NishalCDN25} that could predict a story's impact. Journalists can then adjust their writing to produce more engaging stories that deeply resonate with their local community. Similarly, Warren et al. argue that automated fact-checkers and explainable AI systems should incorporate context-dependent knowledge~\cite{WarrenShowTheWork2025}. They note that this area of research remains nascent.

\subsection{Limitations and Future Work}

In this work, we present findings for local journalists in Germany. We believe that these findings apply to other countries with \add{a} similar media ecosystem. A larger participant pool, spanning multiple countries, would improve our understanding of local journalism workflows and provide additional practical insights into AI-supported reporting systems. In addition, we focused solely on the requirements of non-technical local journalists. We are aware that journalists whose roles resemble those of watchdog or data reporters can propose alternative use cases \add{and data challenges}~\cite{MollerOneSizeFitsSomeJournalisticRole2025}. Our goal is to promote data literacy \add{among} non-technical journalists as digital content \add{continues to} grow.  

Training users and integrating the tools into existing workflows remains challenging, as it would require months of research to yield actionable insights, \add{given that} participants are already occupied with multiple current tasks. Data security and privacy constitute additional constraints, as the majority of journalists reported that their IT departments prohibit them from using unapproved external tools. While interviews \add{help gather} insights, a longitudinal approach that observes participant behavior over time would yield a more comprehensive understanding and a richer view of journalistic practices and the use of AI-supported reporting tools. 

Future research can employ a quantitative approach, involving surveys of journalists in local German newsrooms \add{to complement our work}. We invite the research community and journalists to engage with methods \add{for} promoting data awareness in local journalism. 


\section{Conclusion}

In this paper, we highlight local journalists' challenges when interacting with AI and data, and examine how AI can support them. We found that journalists value the support of automation when used responsibly within ethical boundaries (similarly reported by \cite{DhaeseleerAIDivides2025}). Although they face challenges and skill gaps in working with AI and digital data, our findings suggest that journalists are willing to use AI for \add{certain} editorial processes and mundane tasks. The AI-supported reporting would allow local journalists to focus on in-depth research when writing stories, connecting even more with people and local communities. 

With this work, we provide guidelines and recommendations on how to design journalistic tools having local reporters \add{at} the center. Inspired by the ``Washington Post'' headline, \textit{``Democracy dies in darknes\add{s}''}~\cite{enwiki:1305158484}, we call for more support towards local media, especially as local news deserts are expanding rapidly~\cite{rudolfaugstein2024wuestenradar, medill2024localnewsdeserts}. If local news sites sustain their trustworthiness, they can benefit in an internet filled with fake AI content~\cite{Scire2025_TrustedNewsSitesAI}. Promoting data literacy among journalists improves their ability to \add{work with} data. As \add{a} result, they can use AI and automation responsibly, sustaining meaningful and ethical reporting. In sum, AI-supported reporting can be a handy tool~\cite{NetzwerkRecherche_2025_Positionspapier_KI}, and therefore, we invite researchers and journalists to experiment with such technologies and to shape news reporting that uphold\add{s} the values of a democratic society

\begin{acks}
\revise{We want to thank all the journalists who took the time to participate in this study and share their thoughts. We would also like to thank our colleagues Jasmin Baake and Artur Solomonik for supporting us with some of the interviews conducted in German. We extend our gratitude to the anonymous reviewers for their constructive and thoughtful feedback, which significantly improved the paper.}
\end{acks}



\bibliographystyle{ACM-Reference-Format}
\bibliography{bibliography}

\appendix

\end{document}